\documentclass[a4paper,11pt]{article}
\usepackage{jcappub} 

\usepackage{orcidlink}
\usepackage{aas_macros}
\usepackage{xspace}
\usepackage{graphicx}
\usepackage{subcaption}
\usepackage{amsmath}
\usepackage{multirow}
\usepackage{booktabs}

\newcommand{\td}{\delta}

\newcommand{\kvperp}{\mathbf{k}_{\perp}}

\newcommand{\nvec}{\Hat{\mathbf{n}}}

\newcommand{\inv}{^{\raisebox{.2ex}{$\scriptscriptstyle-1$}}}
\newcommand*\diff{\mathop{}\!\mathrm{d}}

\newcommand*{\eg}{\emph{e.g.}\@\xspace}
\newcommand*{\ie}{\emph{i.e.}\@\xspace}

\newcommand{\Planck}{\emph{Planck}\xspace}
\newcommand{\Abacus}{\texttt{AbacusSummit}\xspace}
\newcommand{\AbacusSummit}{\texttt{AbacusSummit}\xspace}

\newcommand{\Lya}{Ly-$\alpha$\xspace}

\newcommand{\sig}{$\sigma_8$\xspace}
\newcommand{\Sig}{$S_8$\xspace}

\newcommand{\hinvMpc}{\,h\, {\rm Mpc}^{-1}\xspace}
\newcommand{\hinvGpc}{\,h\, {\rm Gpc}^{-1}\xspace}

\def\ltsima{$\; \buildrel < \over \sim \;$}
\def\gtsima{$\; \buildrel > \over \sim \;$}
\def\simlt{\lower.5ex\hbox{\ltsima}}
\def\simgt{\lower.5ex\hbox{\gtsima}}

\providecommand{\sorthelp}[1]{}
\usepackage{booktabs}

\title{Cosmology from Planck CMB Lensing and DESI DR1 Quasar Tomography}

\emailAdd{rbelsunce@lbl.gov, akrolews@uwaterloo.ca}
\affiliation{Remaining Author Affiliations are in Appendix~\ref{sec:affiliations}}

\author[1]{{R.~de Belsunce}\orcidlink{0000-0003-3660-4028},}
\author[2,3,4,*]{{A.~Krolewski}\orcidlink{0000-0003-2183-7021},}
\author[1]{{E.~Chaussidon}\orcidlink{0000-0001-8996-4874},}
\author[5,1]{{S.~Ferraro}\orcidlink{0000-0003-4992-7854},}
\author[1]{{G.~Farren}\orcidlink{0000-0001-5704-1127},}
\author[6,1,5]{{B.~Hadzhiyska}\orcidlink{0000-0002-2312-3121},}
\author[1]{{A.~Tamone},}
\author[2]{{S.~Chiarenza}\orcidlink{0009-0003-6369-9904},}
\author[5]{{N.~Sailer}\orcidlink{0000-0002-5333-8983},}
\author[7]{{C.~Ravoux}\orcidlink{0000-0002-3500-6635},}
\author[1]{{J.~Aguilar},}
\author[8]{{S.~Ahlen}\orcidlink{0000-0001-6098-7247},}
\author[9,10]{{D.~Bianchi}\orcidlink{0000-0001-9712-0006},}
\author[11]{{D.~Brooks},}
\author[1]{{T.~Claybaugh},}
\author[1]{{A.~Cuceu}\orcidlink{0000-0002-2169-0595},}
\author[12]{{A.~de la Macorra}\orcidlink{0000-0002-1769-1640},}
\author[13,14]{{J.~Della~Costa}\orcidlink{0000-0003-0928-2000},}
\author[15,16]{{Biprateep~Dey}\orcidlink{0000-0002-5665-7912},}
\author[11]{{P.~Doel},}
\author[17]{{A.~Font-Ribera}\orcidlink{0000-0002-3033-7312},}
\author[18,19]{{J.~E.~Forero-Romero}\orcidlink{0000-0002-2890-3725},}
\author[20,21,22]{{E.~Gaztañaga},}
\author[1,23]{{S.~Gontcho A Gontcho}\orcidlink{0000-0003-3142-233X},}
\author[24]{{G.~Gutierrez},}
\author[1]{{J.~Guy}\orcidlink{0000-0001-9822-6793},}
\author[25,26]{{H.~K.~Herrera-Alcantar}\orcidlink{0000-0002-9136-9609},}
\author[27,28,29]{{K.~Honscheid}\orcidlink{0000-0002-6550-2023},}
\author[30]{{M.~Ishak}\orcidlink{0000-0002-6024-466X},}
\author[14]{{R.~Joyce}\orcidlink{0000-0003-0201-5241},}
\author[14]{{S.~Juneau}\orcidlink{0000-0002-0000-2394},}
\author[31]{{R.~Kehoe},}
\author[32]{{D.~Kirkby}\orcidlink{0000-0002-8828-5463},}
\author[1]{{T.~Kisner}\orcidlink{0000-0003-3510-7134},}
\author[1]{{A.~Kremin}\orcidlink{0000-0001-6356-7424},}
\author[11]{{O.~Lahav},}
\author[1]{{A.~Lambert},}
\author[33]{{C.~Lamman}\orcidlink{0000-0002-6731-9329},}
\author[1]{{M.~Landriau}\orcidlink{0000-0003-1838-8528},}
\author[34]{{L.~Le~Guillou}\orcidlink{0000-0001-7178-8868},}
\author[1]{{M.~E.~Levi}\orcidlink{0000-0003-1887-1018},}
\author[35,17]{{M.~Manera}\orcidlink{0000-0003-4962-8934},}
\author[27,36,29]{{P.~Martini}\orcidlink{0000-0002-4279-4182},}
\author[14]{{A.~Meisner}\orcidlink{0000-0002-1125-7384},}
\author[37,17]{{R.~Miquel},}
\author[21]{{S.~Nadathur}\orcidlink{0000-0001-9070-3102},}
\author[38,39]{{G.~Niz}\orcidlink{0000-0002-1544-8946},}
\author[26,1]{{N.~Palanque-Delabrouille}\orcidlink{0000-0003-3188-784X},}
\author[2,3,4]{{W.~J.~Percival}\orcidlink{0000-0002-0644-5727},}
\author[40]{{F.~Prada}\orcidlink{0000-0001-7145-8674},}
\author[41]{{I.~P\'erez-R\`afols}\orcidlink{0000-0001-6979-0125},}
\author[27,36,29]{{A.~J.~Ross}\orcidlink{0000-0002-7522-9083},}
\author[42]{{G.~Rossi},}
\author[43]{{E.~Sanchez}\orcidlink{0000-0002-9646-8198},}
\author[1]{{D.~Schlegel},}
\author[44,45]{{M.~Schubnell},}
\author[46]{{H.~Seo}\orcidlink{0000-0002-6588-3508},}
\author[1]{{J.~Silber}\orcidlink{0000-0002-3461-0320},}
\author[14]{{D.~Sprayberry},}
\author[45]{{G.~Tarl\'{e}}\orcidlink{0000-0003-1704-0781},}
\author[14]{{B.~A.~Weaver},}
\author[1]{{R.~Zhou}\orcidlink{0000-0001-5381-4372},}
\author[47]{{H.~Zou}\orcidlink{0000-0002-6684-3997},}

\affiliation{
\noindent \hangindent=.5cm $^{1}${Lawrence Berkeley National Laboratory, 1 Cyclotron Road, Berkeley, CA 94720, USA}\\ 
\noindent \hangindent=.5cm $^{2}${Waterloo Centre for Astrophysics, University of Waterloo, 200 University Ave W, Waterloo, ON N2L 3G1, Canada}\\
\noindent \hangindent=.5cm $^{3}${Department of Physics and Astronomy, University of Waterloo, 200 University Ave W, Waterloo, ON N2L 3G1, Canada}\\
\noindent \hangindent=.5cm $^{4}${Perimeter Institute for Theoretical Physics, 31 Caroline St. North, Waterloo, ON N2L 2Y5, Canada}\\
\noindent \hangindent=.5cm $^{5}${University of California, Berkeley, 110 Sproul Hall \#5800 Berkeley, CA 94720, USA}\\ 
\noindent \hangindent=.5cm $^{6}${Institute of Astronomy, University of Cambridge, Madingley Road, Cambridge CB3 0HA, UK}\\
\noindent \hangindent=.5cm $^{\ast}${ CITA National Fellow} \\ \vspace{-2mm}
}

\abstract{We present a measurement of the amplitude of matter fluctuations over the redshift range $0.8 \leq z \leq 3.5$ from the cross correlation of over 1.2 million spectroscopic quasars selected by the Dark Energy Spectroscopic Instrument (DESI) across 7,200 deg$^2$ ($\sim 170$ quasars$/\mathrm{deg}^2$) and \Planck PR4 (NPIPE) cosmic microwave background (CMB) lensing maps. We perform a tomographic measurement in three bins centered at effective redshifts $z=1.44,\, 2.27$ and 2.75, which have ample overlap with the CMB lensing kernel.
From a joint fit using the angular clustering of all three redshift bins (auto and cross-spectra), and including an $\Omega_m$ prior from 
DESI DR1 baryon acoustic oscillations
to break the $\Omega_m-\sigma_8$ degeneracy,
we constrain the amplitude of matter fluctuations in the matter-dominated regime to be $\sigma_8=0.929^{+0.059}_{-0.074}$ and $S_8\equiv \sigma_8(\Omega_m/0.3)^{0.5} = 0.922^{+0.059}_{-0.073}$.  We provide a growth of structure measurement with the largest spectroscopic quasar sample to date at high redshift, which is $\sim 1.5\sigma$ higher than predictions from $\Lambda$CDM fits to measurements of the primary CMB from \Planck PR4. 
The cross-correlation between PR4 lensing maps and DESI DR1 quasars is detected with a signal-to-noise ratio of $21.7$ and the quasar auto-correlation at $27.2$ for the joint analysis of all redshift bins.  
We combine our measurement with the CMB lensing auto-spectrum from the ground-based Atacama Cosmology Telescope (ACT DR6) and \Planck PR4 to perform a sound-horizon-free measurement of the Hubble constant, yielding $H_0=69.1^{+2.2}_{-2.6}\,\mathrm{km}\,\mathrm{s}^{-1}\mathrm{Mpc}^{-1}$ through its sensitivity to the matter-radiation equality scale.
}

\begin{document}
\maketitle
\flushbottom

\section{Introduction} \label{sec:intro}
The formation of large-scale structure (LSS) through gravitational amplification of density perturbations, as probed by cosmic microwave background (CMB) experiments \cite{Planck2018,Hinshaw2013, SPT2021, Madhavacheril2023} and large spectroscopic surveys \cite{Alam2017, eBOSS2021}, is a prediction of our standard, $\Lambda$ cold dark matter ($\Lambda$CDM), model of cosmology. Conditioned on CMB data, $\Lambda$CDM predicts with exquisite precision the amplitude of matter clustering, $\sigma_8$, (e.g.,~\cite{Planck2018,Planck_lensing:2020}). Gravitational lensing of CMB photons along the line-of-sight probes the gravitational potential of LSS. It traces the projected matter density from present-day to the surface of last scattering with particular sensitivity to the redshift range $1 \simlt z \simlt 5$ (see, e.g.,~\cite{Lewis2006}, for a review). 

Cross-correlation measurements probe the gravitational potential, $\phi$, allowing constraints on the amplitude of the late-time power spectrum, $\sigma_8$. They can also provide power spectrum shape information, parameterized by $\Omega_{\rm m}h$,
from the large-scale power spectrum turnover set by the horizon size at matter-radiation equality.\footnote{The present-day energy density is denoted by $\Omega_m$, $\sigma_8$ is the r.m.s. linear amplitude of matter fluctuations extrapolated to present day in spheres of radius $8\hinvMpc$, and $h$ is the Hubble constant in units $100\, \mathrm{km\, s\inv \, Mpc\inv)}$.} CMB lensing cross-correlations require overlap with tracers at a wide range of redshifts to benefit from the broad distribution of the CMB lensing kernel. The joint analysis of the auto power spectra of a galaxy (or quasar) sample and its cross-correlation with lensing convergence maps allows to break the degeneracy between bias and the growth of structure amplitude. So-called CMB lensing tomography is becoming increasingly powerful as full-sky galaxy samples across a wide range of redshifts are available from the ongoing Dark Energy Spectroscopic Instrument \citep[DESI;][]{2016arXiv161100036D}, Euclid \cite{Euclid20} and LSST \cite{ivezic2019lsst} surveys.

We can also measure cosmological parameters from the weak lensing of distant galaxies, known as cosmic shear.
In general, weak lensing analyses are most sensitive\footnote{Note that the optimal exponent on $\Omega_m$ changes with the lensing source redshift, and is thus slightly different for CMB lensing auto- and cross-correlations than for cosmic shear \cite{Madhavacheril2023,ACT:2023oei}.} to the degenerate combination $S_8\equiv \sigma_8(\Omega_m/0.3)^{0.5}$.
Some cosmic shear and galaxy weak lensing analyses have found a mildly low value of $S_8$ relative to what is expected from the \Planck best-fit $\Lambda$CDM cosmology (see, e.g.,~\cite{Abdalla:2022yfr, 2024arXiv240612106E} for reviews; though see also \cite{KiDSDES,Wright25,Stolzner25} for consistent results with the primary CMB).
Other analyses find mild tensions at the $\sim 2-3 \sigma$ level, \eg peculiar velocities from Supernovae \cite{Stahl:2021mat} or full-shape analyses of the BOSS DR12 galaxy power spectra in the framework of the effective field theory of large-scale structure
\cite{Ivanov:2019pdj,DAmico:2019fhj,Colas:2019ret,Chen:2021wdi,Ivanov:2021fbu} (though see \cite{DESI:2024hhd} for a galaxy power spectrum analysis giving $S_8$ consistent with the CMB). Likewise, some analyses of CMB lensing cross-correlations have also given low values of $S_8$
\cite{Kitanidis:2020xno,Hang:2020gwn,White:2021yvw,Krolewski2021_unw,DES:2022xxr,DES:2022urg,Robertson:2020xom,ACT:2023ipp,Garcia-Garcia:2021unp,Chen:2022jzq,Chen:2024vuf}, but others have yielded consistency with the primary CMB \citep{Qu:2024sfu, 2024arXiv240704607S, 2024arXiv240704606K,ACT:2023oei,ACT:2023dou, Farren:2024rla}.

We present a measurement of $S_8$ using a spectroscopically confirmed quasar sample in the redshift range $0.8 \leq z \leq 3.5$, providing a measurement of the amplitude of structure in the matter-dominated regime.
Spectroscopy provides two key advantages: eliminating error from uncertainty in the redshift distribution, and removing interlopers such as stars and galaxies. This allows us to construct a sample that is higher-density than the high-purity photometric \textit{Quaia} sample of \cite{Alonso23,Storey-Fisher:2023gca} and better characterized and with higher purity than the DESI quasar target sample analyzed in \cite{Krolewski24}.
We detect the cross-correlation between PR4 lensing maps and DESI DR1 with a signal-to-noise ratio of $21.7$ 
and the quasar auto-correlation at $27.2$ for the joint analysis of all redshift bins with the  improvement, compared to previous analyses, stemming from the large number of quasars, the increased sky overlap between the quasar sample and the \Planck lensing map and a higher signal-to-noise ratio in the convergence mass maps. With improved control of large-scale systematics due to the spectroscopic quasar maps, we also present a companion analysis constraining primordial non-Gaussianity from the quasar-CMB lensing cross-correlation in \cite{Chiarenza25}.

This paper is organized as follows: in Sec.~\ref{sec:data} we present the different data sets. In Sec.~\ref{sec:NaMaster} we briefly summarize the angular power spectrum estimator and in Sec.~\ref{sec:theory} we introduce our theoretical model for the angular clustering of quasars and the CMB lensing convergence field and cross-correlations thereof. We extensively test our measurement and likelihood pipeline on Gaussian simulations in Sec.~\ref{sec:Gaussian_simulations} and on realistic \Abacus mocks in Sec.~\ref{sec:Abacus_simulations}. In Sec.~\ref{sec:results} we present our results and conclude in Sec.~\ref{sec:conclusions}.  

\section{Data} \label{sec:data}
In the present paper, we cross-correlate quasars from the DESI survey with lensing mass maps obtained by the \emph{Planck} satellite, namely the PR4 convergence maps. In Sec.~\ref{sec:Planck_data} we briefly present the employed CMB lensing data and summarize the quasar sample in Sec.~\ref{sec:DESI_data}. 

\begin{figure}[t!]
    \centering
    \begin{subfigure}[b]{0.3\textwidth}
        \centering
        \includegraphics[width=\textwidth]{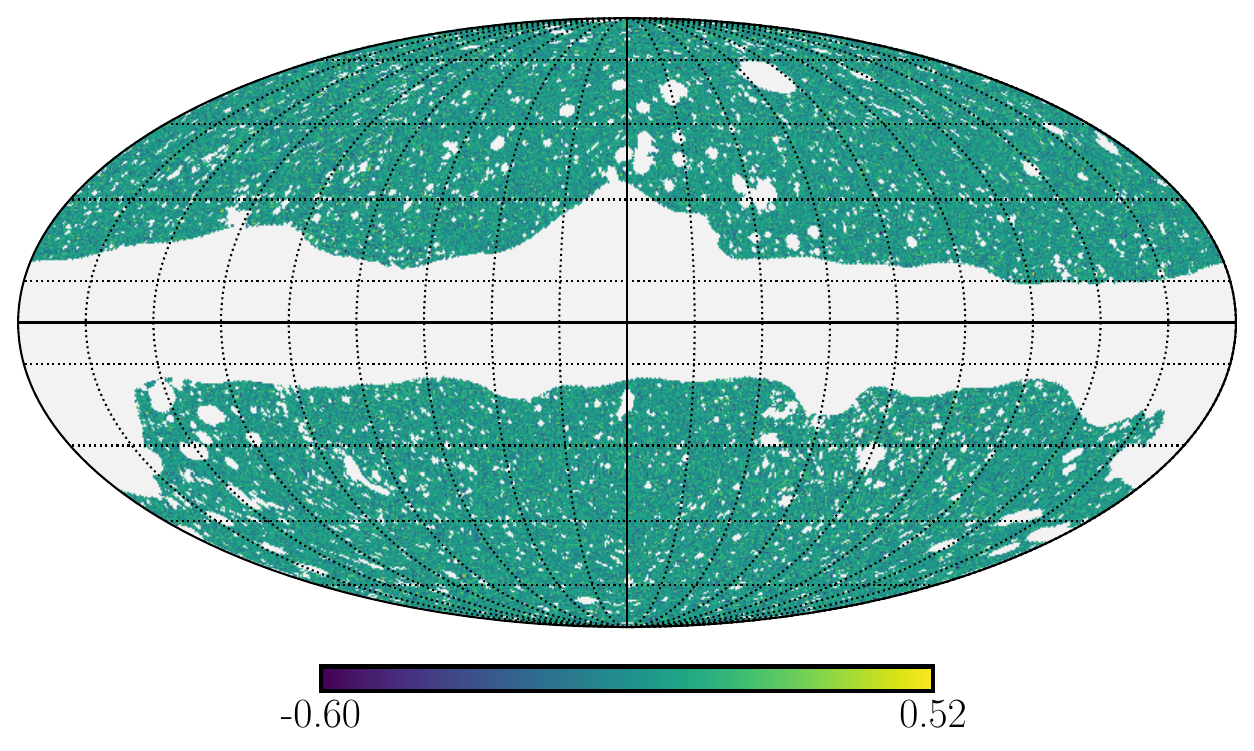}
        \caption{\Planck PR4 lensing map.}
        \label{fig:fig1}
    \end{subfigure}
    \hfill
    \begin{subfigure}[b]{0.3\textwidth}
        \centering
        \includegraphics[width=\textwidth]{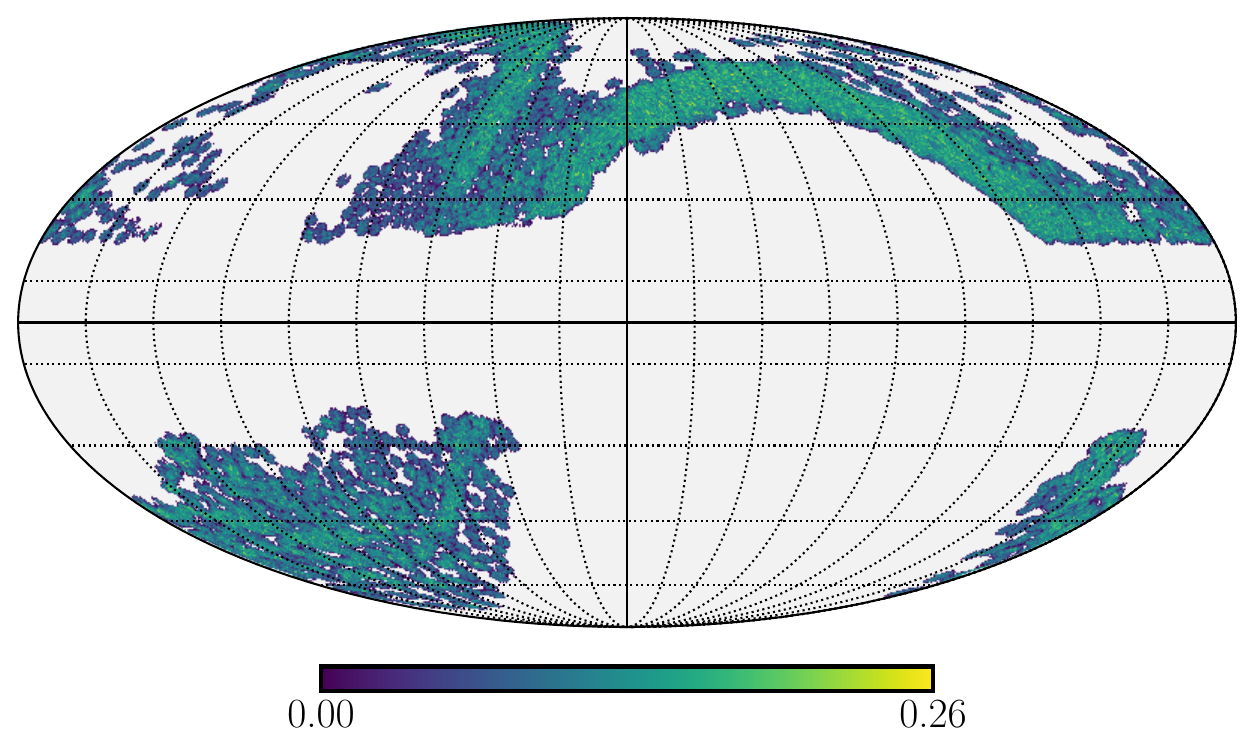}
        \caption{DESI DR1 quasar sample.}
        \label{fig:fig2}
    \end{subfigure}
    \hfill
    \begin{subfigure}[b]{0.3\textwidth}
        \centering
        \includegraphics[width=\textwidth]{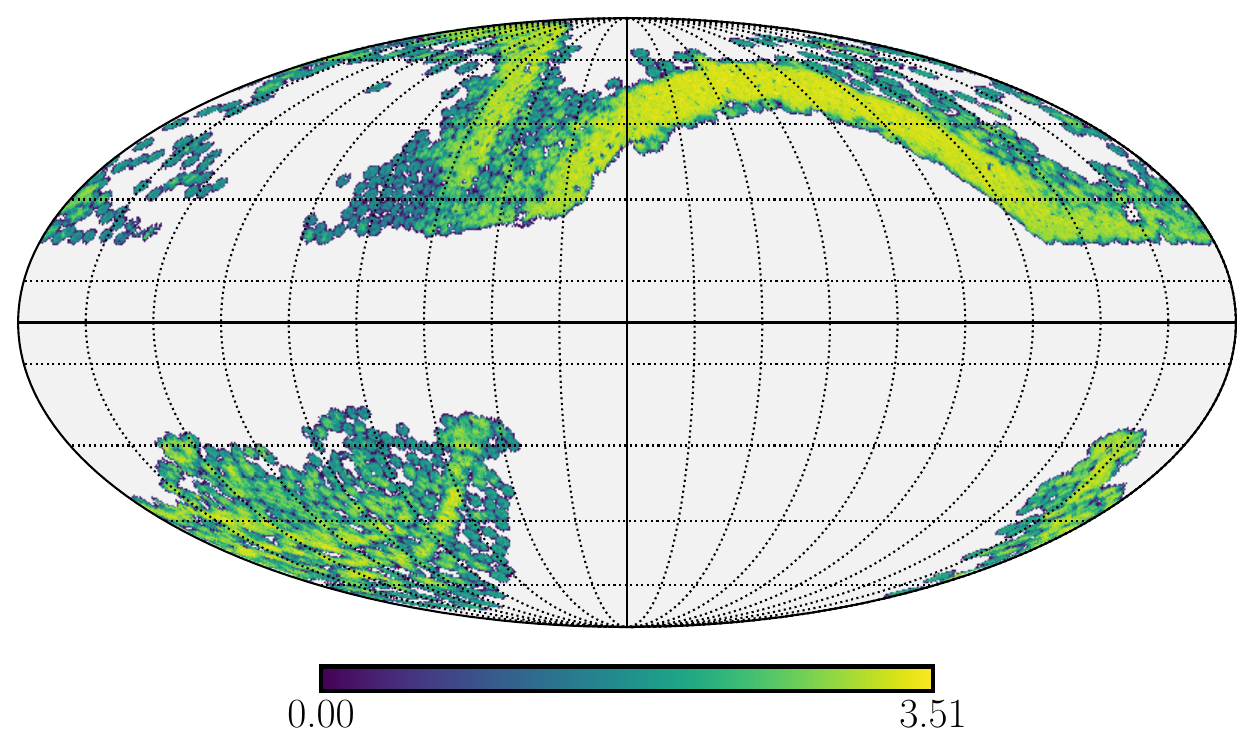}
        \caption{Completeness mask.}
        \label{fig:fig3}
    \end{subfigure}
    \caption{Data used for the cross-correlation. The \Planck PR4 lensing convergence map, $\kappa$, with the corresponding smoothed mask (Gaussian filtered with a $1^{\circ}$ FWHM) is shown in panel (a). The lensing map covers 67.1\% of the sky. In (b) we show the DESI DR1 quasar number counts map for the entire data sample covering all three redshift bins $0.8 \leq z \leq 3.5$ with the corresponding completeness mask in (c). The data covers 20.3\% of the sky with an overlapping sky region between \Planck PR4 and DESI DR1 given in Tab.~\ref{tab:DESI_DR1_data}. For easier readability we show the maps using \texttt{HealPix} resolution of $N_{\rm side}=128$, but the computations use $N_{\rm side}=2048$.
    }
    \label{fig:data}
\end{figure}

\subsection{\textit{Planck} CMB lensing}\label{sec:Planck_data}
CMB photons are gravitationally lensed by intervening matter, creating a distortion in the temperature and polarization fields. The distortions are used to reconstruct the lensing potential \citep[see, e.g.,~][for a review]{Lewis2006}. We use the \Planck PR4 \cite{2022JCAP...09..039C} CMB lensing convergence, $\kappa$, maps\footnote{Publicly available at \url{https://pla.esac.esa.int} and \url{https://github.com/carronj/planck_PR4_lensing}.}. First, we resample the available spherical harmonic coefficients, $\kappa_{\ell m}$, of the minimum variance (MV) maps to $N_{\rm side}=2048$ using the HEALPix convention. For PR4 the NPIPE processing pipeline \cite{2020A&A...643A..42P} applies a more optimal global minimum variance estimator (compared to its predecessor PR3) to the reconstruction. The resulting PR4 CMB lensing map is signal dominated up to $\ell \sim 70$ compared to $\ell\sim 40$ for PR3. This increase in signal-to-noise by $\sim 20\%$ stems from the improved NPIPE processing pipeline together with including an additional $\sim 8\%$ of CMB data (relative to \Planck PR3) from satellite re-pointing and an optimal estimator \cite{Maniyar:2021msb} and anisotropic filtering scheme \cite{Carron:2022eyg}. For our analysis, we apodize the provided binary CMB lensing mask with a 30 arcmin `C2' filter using the \texttt{NaMaster} algorithm \cite{Alonso:2018jzx} (see Sec.~\ref{sec:NaMaster} for more details on angular power spectrum estimation). Additionally, we use the effective reconstruction noise curve $N_{\ell}^{\kappa \kappa}$ provided by the \Planck collaboration when computing the analytic Gaussian covariance matrices introduced in Sec.~\ref{sec:NaMaster}. The sky coverage\footnote{The quantity $f_{\rm sky}$ denotes the fraction of unmasked pixels, $f_{\rm sky} = \sum_{i=1}^{N_{\rm pix}}x_i/N_{\rm pix}^{\rm tot}$. $N_{\rm pix}^{\rm tot}=12 \cdot N_{\rm side}^2 \approx 5\cdot 10^7$ denotes all the pixels in the map while $x_i$ is the value of the mask of pixel $i$.} of the \Planck lensing maps is 67.1\% and the overlap with the tomographic DESI DR1 quasar catalog is tabulated in Table~\ref{tab:DESI_DR1_data}.

\subsection{DESI DR1 quasar sample} \label{sec:DESI_data}
We use 1,223,391 quasars with $0.8 \leq z \leq 3.5$ from the quasar sample \cite{Chaussidon2023} of the first DESI DR1 data release \cite{DESI:2025fxa}. DESI is a highly-multiplexed spectroscopic surveyor operating at the 4-meter Mayall telescope at Kitt Peak National Observatory \cite{DESI:2022}, with the capability of observing 5000 spectra at once \cite{2016arXiv161100037D,2023AJ....165....9S,2023arXiv230606310M,Poppett24, 2023AJ....166..259S}. DESI's five-year survey covers up to 15,000 deg$^2$ and is designed to provide transformative constraints on dark energy via baryon acoustic oscillations and redshift space distortions measured by galaxies and quasars at $0 \lesssim z \lesssim 3.5$ \cite{2013arXiv1308.0847L,DESI:2016}.
After the successful completion of survey validation \cite{DESI:2023dwi} and the early data release \cite{DESI:2023ytc}, the first year of DESI data has already provided
competitive measurements of BAO \cite{DESI:2024mwx,DESI2024.IV.KP6,DESI2024.III.KP4} and redshift space distortions \cite{DESI:2024hhd} and the leading large-scale structure constraint on primordial non-Gaussianity from the large-scale quasar power spectrum \cite{Chaussidon:2024qni}.

\begin{figure}
    \centering
    \includegraphics[width=\linewidth]{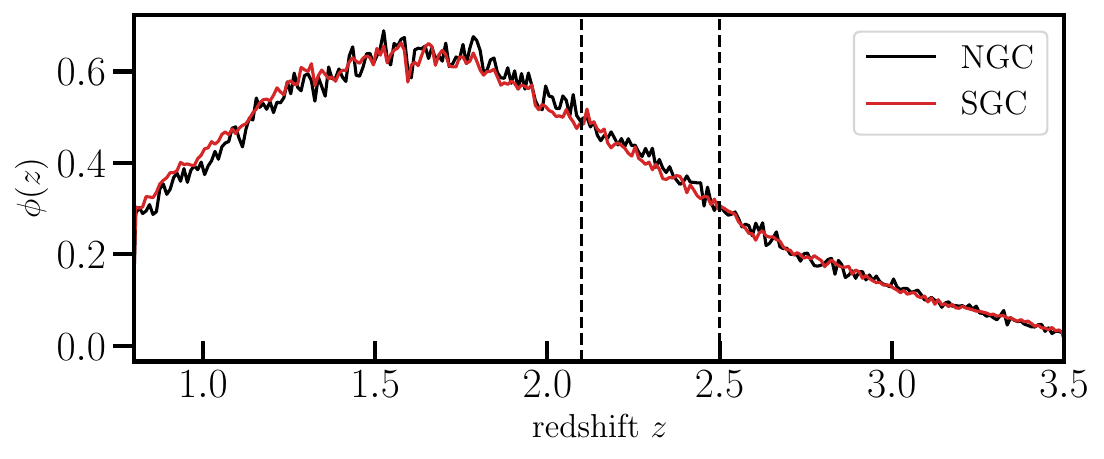}
    \caption{Normalized redshift distribution $\phi(z)$ (with $\int \mathrm{d}z\phi(z)=1$) of the DESI DR1 spectroscopic quasar sample in both regions of the sky: northern galactic cap (NGC) as a black line and southern galactic cap (SGC) as a red line. The vertical black dashed lines demark the boundaries of each non-overlapping redshift bin.}
    \label{fig:DESI_DR1_dNdz_hist}
\end{figure}

An overview of the DESI DR1 quasar sample is given in Table \ref{tab:DESI_DR1_data}, while the redshift distribution is displayed in Figure~\ref{fig:DESI_DR1_dNdz_hist}. We are using the clustering catalogs that are described in \cite{DESI2024II,Ross2024}, and redshift measurements from the spectroscopic pipeline described in \cite{Guy2023,Bailey2023,Brodzeller2023, Anand:2024jrf}.


Due to the different imaging properties of each region, the target selection for the QSOs is adapted for each region \cite{Chaussidon2023}. The discrepancy in either the target selection or imaging quality leads to different mean densities in these different photometric regions, and slightly different redshift distributions as displayed. We use the appropriately-weighted average $n(z)$ for the full sky; on-sky variations in $dN/dz$ at this level do not affect the galaxy-CMB lensing cross-correlation and negligibly affect the galaxy auto-correlation \cite{BaleatoLizancos:2023zpl}.

To correct for observational effects, we weight each quasar by
\begin{equation}
    w_{\rm tot} = w_{\rm sys} \times w_{\rm comp} \times w_{\rm zfail},
\end{equation}
where the weights are described in detail in \cite{Ross2024} correcting for imaging systematics (sys), completeness (comp) and redshift failures (zfail). 
Previous work has found that the fiducial Random Forest imaging systematics weights correction  \cite{2022MNRAS.509.3904C}
can remove power from the CMB lensing cross-correlation on large scales \cite{Krolewski24}. To avoid this, we instead use the linear imaging systematic weights used in the DESI DR1 primordial non-Gaussianity analysis \cite{Chaussidon:2024qni}, which follow the linear regression method from eBOSS \cite{RosseBOSS}. We show in Section~\ref{sec:systematics} that linear regression removes minimal power from $C_\ell^{\kappa g}$ and $C_\ell^{gg}$ on the scales considered.

\begin{table}
    \centering
    \setlength{\tabcolsep}{10pt}
    \begin{tabular}{cccccccc}
    \hline\hline \vspace{-2ex} \\ 
    Label & $z$-range & $N_{\rm qso}$ & shotnoise & $\overline{z}$& $z_{\rm eff}$ &$s_{\mu}$ &$f_{\rm sky}^{\rm PR4\, x\, DR1}$  \\ [0.5ex] 
    \hline \vspace{-2ex} \\ 
    $g_1$& $0.8 \leq z < 2.1$ & 856,831 & $\phantom{1}2.6\cdot 10^{-6}$ & 1.49&1.44& 0.099& 19.8\\ [0.5ex]
    $g_2$& $2.1 \leq z < 2.5$ & 194,754 & $11.2\cdot 10^{-6}$ & 2.28&2.27& 0.185&18.9\\ [0.5ex]
    $g_3$& $2.5 \leq z \leq 3.5$ & 171,806 & $12.7\cdot 10^{-6}$ & 2.85 &2.75&0.244&18.6 \\ [0.5ex] \hline
    \end{tabular}
    \caption{Overview of DESI DR1 quasar samples used for the cross-correlation with \Planck CMB lensing. The total number of quasars used for the cross-correlation analysis from the DESI DR1 sample is 1,223,391. We list all three redshift bins, denoted by $\{g_1,g_2,g_3\}$, with different redshift coverage in the range $0.8 \leq z \leq 3.5$, number of quasars per bin, the mean  $\overline{z}$ and effective redshift $z_{\rm eff}$ (calculated as in \cite{Sailer2024}), the magnification bias $s_{\mu}$, and sky coverage overlap of the DESI DR1 sample with the CMB lensing convergence maps PR4, $f^{\rm PR4xDR1}_{\rm sky}$, in per cent.}
    \label{tab:DESI_DR1_data}
\end{table}

\section{Angular Power Spectrum Estimator}\label{sec:NaMaster}
To make inferences from large data sets such as projected quasar and CMB lensing maps, data compression is required to reduce the size of the data vector to a manageable level and average over stochastic fluctuations. Here, we compress maps into summary statistics, namely the angular power spectra\footnote{We use the following notation: window-convolved power spectra are denoted by $\widehat{C}_{\ell}$ and theory spectra by $C_{\ell}$ (or $C_{\ell}^{\rm theory}$).} of two pixelized maps\footnote{Instead of pixelizing the projected quasar map, one could also directly compute the spherical harmonic transform of discrete tracers (see, e.g., ~\cite{Baleato2024}). On the scales that we consider, we find very good consistency between the pixel-free angular power spectrum and the pixelized angular power spectrum. The pixel-free framework allows for a more robust and accurate treatment of large angular scales, and so we use it in our companion paper constraining primordial non-Gaussianity \cite{Chiarenza25}. We find that the pixelized approach in our work leads to significant excess power in the auto-correlation at $\ell < 30$, but we do not use these scales.}, \ie the pseudo-$C_{\ell}$'s. The lensing convergence maps are given in the standard HEALPix  \cite{Gorski:2004by} pixelization scheme and the quasar distribution is projected in tomographic redshift bin onto the sky using the same pixelization scheme. Our measurement pipeline is based on the \texttt{NaMaster} implementation \citep{Alonso:2018jzx} of the \texttt{MASTER} algorithm \citep{Hivon2002}. 

We compute the angular power spectra of two fields $X$ and $Y$ using a projected quasar map in three redshift bins, denoted by $g_1,\, g_2,\, g_3$, and the lensing convergence map, denoted by $\kappa$. The window-convolved measured pseudo-$C_{\ell}$'s are given by
\begin{equation}
    \widehat{C}_{\ell}^{XY} = \frac{1}{2\ell+1} \sum_{m=-\ell}^{\ell} a_{\ell m}^Xa_{\ell m}^{Y\, \ast}\ ,
\end{equation}
where the $a_{\ell m}^{X,Y}$ are spherical harmonic decompositions of the respective field. To obtain the unbiased and window deconvolved angular power spectrum, we apply the inverse of the mode-coupling matrix $M_{\ell\ell'}$ \cite{Hivon2002}
\begin{equation} \label{eq:Mll}
    \left\langle \widehat{C}_\ell \right\rangle
    = \sum_{\ell'} M_{\ell\ell'} C_{\ell'}\ , \qquad  M_{\ell\ell'} = \frac{2\ell'+1}{4\pi} \sum_{\lambda} (2\lambda+1)
    \begin{pmatrix} \ell & \ell' & \lambda \\ 0&0&0 \end{pmatrix}^2
    W_{\lambda}\,,
\end{equation}
which is computed from the power spectrum of the mask, $W_\lambda$, and multiplied by the Wigner-$3j$ symbols
where $C_{\ell'}$ is the `true' angular power spectrum of the overdensity field. We use Eq.~\eqref{eq:Mll} to deconvolve the quasar mask\footnote{Alternatively, we could also forward model the effect of the mask and apply it to the theory spectra. We have tested on simulations that we recover the same results to better than 0.1\%.}. Note that while the mode-coupling matrix $M_{\ell\ell'}$  is (usually) not invertible, re-binning it into bandpowers under the assumption that the $C_{\ell}$'s are constant within each bin $b$ allows for inversion of the binned mode-coupling matrix (see, e.g., ~\cite{Alonso:2018jzx}). Therefore, we measure the angular power spectra in bandpowers: $C_b = \sum_{b'}M_{bb'}\inv \widehat{C}_{b'}$, starting at $\ell = 5$ and with $\Delta \ell = 20$. However, since the theory $C_{\ell}$'s are not constant in some bandpower $b$, we do the following: First, we window-convolve the theory with $M_{\ell\ell'}$ to obtain $\widehat{C}_{\ell}^{\rm theory}$, bin it in bandpowers and window-deconvolve it with $M_{bb'}$. Theory spectra are generated with the Boltzmann solver \texttt{CAMB}\footnote{Publicly available at \url{https://camb.info/}.} \cite{Lewis:1999bs,Howlett:2012mh}. 

For the present analysis, we use a Gaussian analytic covariance \cite{GarciaGarcia19}, implemented in \texttt{NaMaster}. Therefore, a set of fiducial spectra and noise estimates is required: (i) For the lensing power spectrum $C_{\ell}^{\kappa \kappa}$ we use the publicly available PR4 lensing spectra with the provided lensing reconstruction noise $N_{\ell}^{\kappa \kappa}$; (ii) we use an iterative approach by first using a set of fiducial spectra for the cross-correlation between lensing and the different quasar redshift bins in a fiducial cosmology with $h = 0.67$, $A_s = 2.1\times10^{-9}$, $n_s = 0.96$, $\Omega_c = 0.27$, $\Omega_b = 0.045$, one 0.06 eV massive neutrino, the quasar bias $b(z)$ from \cite{eBOSS:2017ozs}, and higher-order bias parameters set to zero. We fit for cosmological and nuisance parameters with these covariance matrices and use the resulting best-fit spectra as fiducial spectra to compute the covariance matrices. (iii) Further, we include inter-quasar correlations $C_{\ell}^{g_ig_j}$ using the best-fit values of the linear bias parameters and use the resulting best-fit spectra as our final fiducial spectra. These are computed up to $ \ell_{\rm max} = 3 \cdot \texttt{Nside}=6,144$. 


\section{Theoretical Model and Likelihood} \label{sec:theory}
\subsection{Angular clustering} \label{sec:theory_cl}
We follow standard practice and relate the three-dimensional matter power spectrum to the two-dimensional angular power spectrum (see, e.g.,~\cite{Hu2004}) using the Limber approximation\cite{1953ApJ...117..134L,1992ApJ...388..272K}\footnote{Note that we have used the second order Limber approximation $k_{\perp}=(\ell+1/2)/\chi$ to further increase the accuracy to $\mathcal{O}(\ell^{-2})$ \citep{LoVerde2008}. This approximation is accurate at the per-cent level at the analyzed scales $\ell>30$. } within some redshift bin
\begin{align}  \label{eq:Cell_general}
C^{XY}_{\ell} = \int_0^{\chi_{\ast}}\diff \chi\,  \frac{W^X(\chi)W^Y(\chi ) }{\chi^2} 
P_{XY}\left(\kvperp = \frac{\ell+1/2}{\chi}, k_\parallel=0,\chi \right) \ ,
\end{align}
where $P_{XY}$ denotes the equal-time cross-power spectrum of both tracers. The goal of the present paper, is to cross-correlate the DESI DR1 quasar sample $X=q$ (or $g$) and the CMB lensing convergence field $Y=\kappa$. In the following, we present the corresponding projection kernels, $W^{\kappa,\, g}(\chi)$. 

CMB lensing convergence, $\kappa$, is related to the underlying matter density field by setting the lensing source to be at the surface of last scattering \cite{Lewis2006}
\begin{equation}\label{eq:Wk} 
    \kappa(\nvec) = \int_{0}^{\chi_{\ast}} \diff \chi \, W^{\kappa} (\chi) \td_m(\chi \nvec, \chi) \quad W^{\kappa}(\chi) = \frac{3}{2}\frac{\Omega_{m} H_0^2}{ a} \frac{\chi(\chi_{\ast}-\chi)}{\chi_{\ast}}
\end{equation}
where $W^{\kappa}(\chi)$ is the well-known lensing kernel
and $\Omega_{m}$ the present-day total matter density which includes the  neutrino density, $H_0$ the Hubble constant, and $a=1/(1+z)$ the scale factor. The comoving distance to the surface of last scattering is denoted by $\chi_{\ast} = \chi(z_{\ast}\simeq 1100) \simeq 9,400 \hinvMpc$.

The quasar (or galaxy) projection kernel encodes the quasar bias and the redshift distribution of the sample
\begin{equation}
    W^{g}(\chi) =b(z(\chi)) \frac{\diff N}{\diff \chi}\, ,
\end{equation}
which is normalized such that $\int \diff z\, \diff N/\diff z=1$. We assume a fixed redshift dependence for $b(z)$, following \cite{Chaussidon:2024qni}, and freeing the linear bias means that we freely scale $b(z)$.
We also include contributions stemming from lensing magnification of background sources, called magnification bias $s_{\mu}$, which contribute at the five per-cent level \cite{Krolewski2021_unw}:
\begin{equation}
    W^{\mu,i}(\chi) = (5s_{\mu}-2)\frac{3}{2}\Omega_mH_0^2(1+z)g_i(\chi)\, \qquad g_i(\chi)=\int_{\chi}^{\chi_{\ast}} \diff \chi' \frac{\chi(\chi'-\chi)}{\chi'}H(z')\frac{\diff N_i}{\diff z'}\, ,
\end{equation}
where $s_{\mu}$ is the response of the quasar number density to perturbations in the magnitude. In summary, our model for the angular clustering is given by the density-density and density-lensing and magnification bias contributions $\{C_{\ell}^{\kappa q}=C_{\ell}^{\kappa q}+C_{\ell}^{\kappa \mu},\, C_{\ell}^{q q}=C_{\ell}^{q q}+2C_{\ell}^{q \mu}+C_{\ell}^{\mu \mu},\, C_{\ell}^{\kappa \kappa} \}$. Note that we include the effect of redshift-space distortions (RSD) in the kernel of the projected quasar density following Eq.~9 from \cite{2019MNRAS.489.3385T}. 
\begin{align}
\begin{split}
    W^g(\chi) \rightarrow W^g(\chi) \, &+ \, \frac{2\ell^2 + 2\ell - 1}{(2 \ell - 1) (2 \ell + 3)} W^{f,i}(\chi) \, - \,  \frac{\ell (\ell - 1)}{(2\ell - 1) \sqrt{(2 \ell - 3) (2 \ell+1)}} W^{f,i}\left(\frac{2\ell-3}{2\ell+1}\chi\right) \\ 
    &- \, \frac{ (\ell + 1)(\ell + 2)}{(2\ell + 3)\sqrt{(2 \ell + 1)(2\ell + 5})}W^{f,i}\left(\frac{2\ell + 5}{2\ell+1}\chi\right)
\end{split}
\end{align}
with the projection kernel
\begin{equation}
W^{f,i} = \frac{dN}{d\chi} f(z)
\label{eqn:rsd_kernel}
\end{equation}
where $f$ is the growth rate, $d\ln{D}/d\ln a$.
The fractional contribution of the RSD term is largest on large scales, and it contributes at the 0.1\% level to $C_{\ell}^{\kappa g}$  at $\ell_{\textrm{min}} = 45$, and at the 1\% level to $C_{\ell}^{gg}$  at $\ell_{\textrm{min}} = 85$.

\subsection{Non-linear power spectrum modeling} \label{sec:theory_pk}
We model the auto- and cross-correlation using a hybrid approach (see, e.g., ~\cite{Krolewski2021_unw,ACT:2023oei}), combining fits to N-body simulations and Convolutional Lagrangian Effective Field Theory (CLEFT; \cite{2017JCAP...08..009M}). First, we generate a non-linear \texttt{halofit} (HF; \citep{2016MNRAS.459.1468M}) power spectrum with the Boltzmann code \texttt{CAMB}. Then, we add the leading-order non-linear bias terms from one-loop Lagrangian perturbation theory computed with the \texttt{velocileptors} code\footnote{Publicly available at \url{https://github.com/sfschen/velocileptors}.} \cite{Chen:2020fxs}. For completeness, the quasar-CMB lensing cross-spectrum (gm), the quasar auto-spectrum (gg), and the matter spectrum (mm) are modeled as \citep{Krolewski2021_unw}
\begin{align}
P_{gm}(k,z) &= b_{1,E}(z) P_{mm,HF}(k,z) + \frac{b_{2,L}(z)}{2} P_{b_2}(k,z) + \frac{b_{s,L}(z)}{2} P_{b_s}(k,z) \\
P_{gg}(k,z) &= b_{1,E}(z)^2 P_{mm,HF}(k,z) + b_{2,L}(z) P_{b_2}(k,z) + b_{s,L}(z) P_{b_s}(k,z) \\
&+ b_{1,L}(z) b_{2,L}(z) P_{b_1 b_2}(k,z) + b_{1,L}(z) b_{s,L}(z) P_{b_1 b_s}(k,z)  + b_{2,L}(z) b_{s,L}(z) P_{b_2 b_s}(k,z) 
\nonumber \\
&+ b_{2,L}(z)^2 P_{b_2^2}(k,z) + b_{s,L}(z)^2 P_{b_s^2}(k,z) + \text{SN}\nonumber \\
P_{mm}(k,z) &= P_{mm,HF}(k,z) 
\end{align}
where Eulerian (Lagrangian) biases are denoted by E (L) which are related through $b_{1,E}=b_{1,L}+1$ and $P_{mm,HF}$  is the \texttt{halofit} matter power spectrum.
The power spectra $P_X$ with $X \in \{b_1, b_2, b_s,b_1b_2, b_1b_s, b_2^2, b_s^2\}$ are the one-loop auto-spectra in redshift space in Lagrangian perturbation theory (see, e.g., ~\cite{2021JCAP...03..100C}), and SN denotes the shot-noise over which we analytically marginalize. 
We follow \cite{Chen:2022jzq} in our treatment of neutrinos, using $P_{\textrm{cb},\textrm{cb}}^{\textrm{lin}}$ as input for the perturbative terms in the galaxy auto-spectrum, $P_{\textrm{cb},\textrm{m}}^{\textrm{lin}}$
as input for the galaxy cross-spectrum, and $P_{\textrm{mm}}$ (including neutrinos) for the matter auto-spectrum.
Likewise, we use $P_{\textrm{cb},\textrm{cb}}^{\textrm{HF}}$ for the nonlinear galaxy auto-spectrum and
$P_{\textrm{cb},\textrm{m}}^{\textrm{HF}}$ for the nonlinear galaxy-matter cross-spectrum.
In our implementation, we vary the cosmology for \textit{all} computed spectra when fitting data or simulations. For the RSD terms, we use the same (real-space) expansions for the power spectra and the linear-theory RSD kernel from Eq.~\eqref{eqn:rsd_kernel}.
This simplification is justified because the RSD contribute only a small fraction to the total angular power spectrum, and are largest on large, linear scales. We use this approximation because it considerably simplifies our modelling of this small term.

\subsection{Likelihood pipeline and priors}\label{sec:likelihood}
To constrain the cosmological parameters, $\textbf{p}$, for model, $\textbf{m}$, given the quasar and CMB lensing data set, $\mathbf{d}$, we use a multi-variate Gaussian likelihood:
\begin{equation}
-2 \log \mathcal{L}(\mathbf{d}|\mathbf{p}) = [\textbf{m}(\textbf{p}) - \textbf{d}]^T C\inv [\textbf{m}(\textbf{p}) - \textbf{d}] + \log |2\pi C|\,,
\end{equation}
where $C$ is the covariance matrix. Therefore, we sample from the posterior using a Markov chain Monte Carlo (MCMC) as implemented in \texttt{Cobaya} \cite{2021JCAP...05..057T}. Chain convergence is computed using the Gelman-Rubin statistic set to $R-1<0.03$ \footnote{Each run takes approximately 0.5-1 CPU hour on a NERSC AMD Milan CPU.}. To analyze the chains we remove a burn-in phase of 30\% and use the \texttt{getdist} package \cite{Lewis:2019xzd}. In table \ref{tab:priors} we quote the used priors: the first block quotes the sampled cosmological parameters, the second one gives the fixed cosmological parameters to \Planck best-fit parameters \cite{Planck2018} with a prior on $\Omega_mh^3$ which is a proxy for the exquisitely measured $\theta_{\ast}$ from the CMB, \textit{ie.}~a geometric constraint from the angular sound horizon scale at recombination, and the fixed values of the baryon abundance from Big Bang Nucleosynthesis measurements \cite{Schoneberg:2019wmt}. The bottom block gives the priors on the nuisance parameters for each redshift bin $z_i$. The values for the magnification bias and shot noise are given in table \ref{tab:DESI_DR1_data}. We analytically marginalize over the shot noise parameter as it enters the theoretical model linearly. The theory prediction can be split into theory templates $\mathbf{A}$ and $\mathbf{B}$ that depend on a set of model parameters $\mathbf{\theta}$ and a linear dependence on parameters $\mathbf{\phi}_i$,  (see, e.g.,~\cite{2024arXiv240704607S, deBelsunce:2024rvv} for more details).  

\begin{table}
    \centering
    \setlength{\tabcolsep}{10pt}
    \begin{tabular}{llc}
        \hline\hline \vspace{-2ex} \\ 
        Parameter & Description& Prior\\  [0.5ex]
        \hline
        $\Omega_m$ &dark matter density& $\mathcal{U}(0.1, 0.9)$ \\ [0.5ex]
        $\ln(10^{10}A_s)$ & primordial power-spectrum amplitude& $\mathcal{U}(1.0, 5.0)$ \\ [0.5ex]
        \hline \vspace{-2ex} \\ 
        $\Omega_bh^2$ &  baryon abundance \cite{Schoneberg:2019wmt} &$0.02236$ \\ [0.5ex]
        $n_s$ &  spectral index \cite{Planck:2018vyg} &$0.9649$ \\ [0.5ex]
        $\Omega_mh^3$ & proxy for the angular acoustic scale \cite{Krolewski2021_unw} &$0.09633$ \\ [0.5ex]
        $\sum m_{\nu}$ &sum of neutrino masses \cite{Planck:2018vyg} &$0.06$ \\ [0.5ex]
        \hline\hline \vspace{-2ex} \\ 
        $b_1(z_i)$ &linear bias&$\mathcal{U}(-10, 10)$\\ [0.5ex]
        $b_2(z_i)$ &quadratic bias&$\mathcal{U}(-10, 10)$\\ [0.5ex]
        $b_s(z_i)$ &shear bias&$\mathcal{N}(0, 1^2)$\\ [0.5ex]
        $s_{\mu}(z_i)$ &magnification bias&$\mathcal{N}(s_{\mu}(z_i), \left(s_{\mu}(z_i)/3\right)^2)$  \\ [0.5ex]
        $\text{SN}(z_i)$ & shot noise &$\mathcal{N}(\text{SN}(z_i), \left(\text{SN}(z_i)/3\right)^2)$\\ [0.5ex]
        \hline
    \end{tabular}
    \caption{Priors for the parameters used in the likelihood analysis. The first block of two rows are the sampled cosmological parameters. The second block of four rows are fixed cosmological parameters and the remaining parameters in the third block are the sampled nuisance parameters for each redshift bin $z_i$. Uniform priors are denoted by $\mathcal{U}(x_1, x_2)$, and Gaussian priors by $\mathcal{N}(\mu, \sigma)$. The values for the magnification bias and shot noise per redshift bin are given in table \ref{tab:DESI_DR1_data}. For those we choose a relative width of $\sim 33\%$ of the mean value of the measured magnification bias and shot noise, and our results are insensitive to the size of the priors chosen.}
    \label{tab:priors}
\end{table}

\subsection{Scale cuts}
For the present analysis, we impose low- and high-$\ell$ scale cuts. For the auto-correlation we use $\ell_{\rm min}=85$ and for the cross-correlation we use $\ell_{\rm min}=45$. 
This allows us to avoid large-scale modes in the autocorrelation, where we see anomalous cross-correlations between the redshift bins at $\ell < 85$ and would also need to apply a small correction for the angular integral constraint. To be conservative, we also remove large-scale modes in the cross-correlation, although we see no evidence for systematics.

We apply a high-$\ell$ cut based on the values of $k_{\text{max}}\, [\text{Mpc}^{-1}]$, computed at the mean redshift of each bin, $\overline{z}$. This yields an $\ell_{\text{max}}= k_{\text{max}}\chi(\overline{z})$ where we assume a \Planck 2018 fiducial cosmology \citep{Planck2018} to compute the comoving radial distance $\chi(\overline{z})$. For our analysis we use three $\ell_{\text{max}} = (605,\, 805,\, 1205)$ cuts (except for the lowest redshift bin where we only use the first two). Each $\ell_{\text{max}}$ corresponds to a different maximum wavenumber $k_{\text{max}}$ based on the effective redshift of each bin. For the lowest redshift bin each scale cut corresponds to $k_{\text{max}}^{\overline{z}_1}\, [\text{Mpc}^{-1}] = 0.13,\, 0.18$, for the center bin to $k_{\text{max}}^{\overline{z}_2}\, [\text{Mpc}^{-1}] = 0.11,\, 0.14,\, 0.21$ and for the highest redshift bin to $k_{\text{max}}^{\overline{z}_3}\, [\text{Mpc}^{-1}] = 0.09,\, 0.13,\, 0.19$, respectively. Our baseline analysis uses $\ell_{\text{max}} = 605,\, 805,\, 1205$ (corresponding to $k_{\textrm{max}} = 0.14$ Mpc$^{-1}$) and we compare our results using two more conservative scale cuts: (i) $\ell_{\text{max}} = 605,\, 805,\, 805$, and (ii) $\ell_{\text{max}} = 605,\, 605,\, 605$. 
Our fiducial $k_{\textrm{max}}$ cut restricts the analysis to quasi-linear scales and is similar to the scales chosen in the DESI 2024 full-shape analyses or in other CMB lensing cross-correlation analyses.
The resulting data vector for the baseline analysis is of size 282 with 14 degrees of freedom corresponding to $26,\, 56,\, 56$ auto- and $28,\, 58,\, 58$ cross-bandpowers, respectively. The conservative scale cut yields for variant (i) a data vector of size 202 with $26,\, 36,\, 36$ auto- and $28,\, 38,\, 38$ cross-bandpowers, and for variant (ii) a data vector of size 162 with $26,\, 26,\, 26$ auto- and $28,\, 28,\, 28$ cross-bandpowers, respectively.

\subsection{Pixel window function}
We distribute the quasars on a \textsc{HEALPix} grid at $N_{\rm side}=2048$. This results in a convolution of the tracers with a pixel window function, effectively yielding aliasing power. Therefore, the spectra are multiplied by the pixel window $w_{\ell}$ using the routine from \texttt{healpy}'s \texttt{pixwin} function, replacing the theory spectra by $C_{\ell}^{\kappa q} \rightarrow w_{\ell}C_{\ell}^{\kappa q}$ and $C_{\ell}^{qq} \rightarrow w^2_{\ell}C_{\ell}^{q q} + \mathrm{SN}$. Note that (i) the auto-spectra are convolved with the pixel window \textit{prior} to adding the shot noise; and (ii) the effect of $w_{\ell}^2$ is approximately 0.8, 1.4, 3.2\% for our scale cuts $\ell_{\rm max}=605,\, 805,\, 1205$, respectively. To avoid assuming a fiducial value of the shot noise when accounted for at the data level (see, e.g.,~\cite{White:2021yvw, Krolewski2021_unw}), we forward model the pixel window function (similar to \cite{2024arXiv240704607S}) and directly apply it to the theory spectra. 

\subsection{Bias evolution}
We model the higher-order biases using bias relations measured from clustering of halos in $N$-body simulations from \cite{Abidi:2018eyd}. Therefore, we use the bias values from figure~8 of~\cite{Abidi:2018eyd} together with the best-fit linear bias relation from DESI DR1 QSOs \cite{Chaussidon:2024qni}
\begin{equation} \label{eq:bias_evolution}
    b_1(z) = a  (1 + z)^2 + b\,, \quad \text{with} \quad a, b =0.237 \pm 0.010, 0.771 \pm 0.070\,.
\end{equation}


\section{Consistency tests on Simulations} \label{sec:simulations}
Before presenting the main results of this paper, we validate our pipeline on two sets of simulations. First, we use Poisson-sampled Gaussian mocks from input power spectra representative of the DESI quasars and CMB lensing to validate the window function treatment and, second, on  \AbacusSummit lightcone mocks that match the DESI DR1 quasar number density to test the validity of the theoretical model. The parameter recovery results for Gaussian simulations are presented in Sec.~\ref{sec:Gaussian_simulations} and for the \AbacusSummit mocks in Sec.~\ref{sec:Abacus_simulations}. 

\subsection{Likelihood validation on Gaussian mocks} \label{sec:Gaussian_simulations}
The first series of tests of the presented theory pipeline and (well established) angular power spectrum estimator presented in
Secs.~\ref{sec:NaMaster} and \ref{sec:theory} is on Poisson-sampled Gaussian mocks of a DESI-like quasar sample. Therefore, we generate a Gaussian realization of a theory lensing power spectrum, $C_{\ell}^{\kappa\kappa}$ and noise curve taken from \Planck PR3.\footnote{Publicly available at \url{https://pla.esac.esa.int}.} The quasar map is obtained by computing Gaussian realizations\footnote{This can be done using the \texttt{synfast} routine from \texttt{healpy}.} of (i) the ratio of the squared cross-spectrum, $\left(C_{\ell}^{\kappa g}\right)^2$, to the lensing spectrum, $C_{\ell}^{\kappa \kappa}$; and (ii) the difference between a fiducial auto spectrum of the quasars (appropriate for the $0.8 < z < 2.1$ bin) and the ratio from step (i). We assume a quasar density of $114\ \mathrm{deg}^{-2}$ which corresponds to a shot noise of $2.67\cdot 10^{-6}$. To obtain the quasar number count maps, we use a map with randoms from the DESI pipeline. 
We use a fiducial $\Lambda$CDM linear power spectrum as theory input \citep{Planck2018} and sample the posterior using MCMC.

\begin{figure}
    \centering
    \includegraphics[width=1\linewidth]{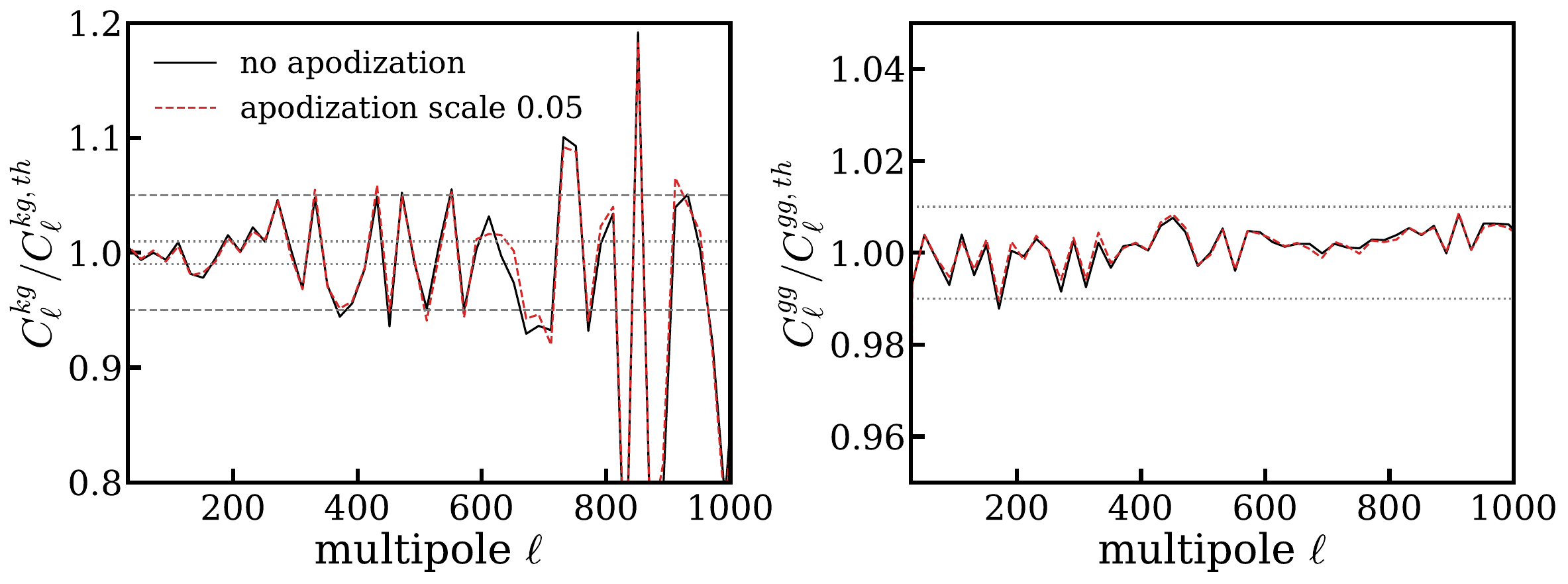}
    \caption{Transfer function of the window deconvolution for the cross (left) and auto-spectra, comparing both the case where the galaxy mask is not apodized (black) and where it is apodized by 0.05$^\circ$  using the ``C2'' filter from \textsc{NaMaster}. We show the ratio of the mean measured spectra divided by the input theory. To guide the eye we include a horizontal dashed (dotted) line illustrating deviations at the 5\% (1\%) level. Both spectra show that the mask deconvolution does not introduce any biases: the cross-correlation is noisy but accurate up to $\ell\sim 805$ at the 8\% level and the auto spectrum at the per-cent level at all scales.}
    \label{fig:transfer_function}
\end{figure}

In figure~\ref{fig:transfer_function} we show the accuracy of the mask deconvolution procedure, introduced in Sec.~\ref{sec:NaMaster}. The cross-spectrum is unbiased but noisy above $\ell\sim 800$ and the auto-spectrum is accurate at the per-cent level at all scales. For our cosmological analysis, we do not include a simulation-based transfer function to correct the measured spectra since these introduce noise into the parameter sampling step.\footnote{We have verified that including a transfer function does not affect the main conclusion of this paper.} We apply a smoothing of 0.05$^\circ$ using the ``C2'' filter implemented in \textsc{NaMaster} to the quasar mask by default. 


\begin{table}
\centering
\setlength{\tabcolsep}{2pt}
\begin{tabular}{lcccccc}
\hline\hline \vspace{-2ex} \\ 
 & $\Omega_m$ & $\frac{\Delta\Omega_m}{\sigma_{\Omega_m}}$ & $\sigma_8$ & $\frac{\Delta\sigma_8}{\sigma_{\sigma_8}}$ & $S_8$ & $\frac{\Delta S_8}{\sigma_{S_8}}$ \\ [1ex]
\hline \vspace{-2ex} \\ 
\multicolumn{7}{l}{\textbf{Gaussian simulations}}\\[1ex]
Theory & $0.310^{+0.056}_{-0.062}[0.313]$ & 0.11& $0.836^{+0.0613}_{-0.053}[0.830]$ & 0.13 & $0.849 ^{+0.076}_{-0.077}[0.848]$ & 0.02  \\[1ex]
$100$ mocks & $0.336\pm 0.062[0.335]$ & 0.31& $0.822\pm 0.063[0.830]$ & 0.10 & $0.876\pm 0.072 [0.870]$ & 0.35  \\[1ex]
\hline \vspace{-2ex} \\ 
\multicolumn{7}{l}{\textbf{\texttt{AbacusSummit} no downsampling \& full-sky -- DESI $0.8 < z < 2.1$ error}}\\[1ex]
\texttt{CLEFT}& 
$0.346^{+0.037}_{-0.054}[0.337]$ & 0.47& $0.819^{+0.050}_{-0.054}[0.808]$ & 0.00 & $0.875^{+0.054}_{-0.065}[0.856]$ & 0.31 \\[1ex] 
$ +\,  \Omega_m^{\rm DR1}$ prior & 
$0.316^{+0.015}_{-0.015}[0.315]$ & 0.03& $0.832^{+0.046}_{-0.046}[0.830]$ & 0.20 & $0.853^{+0.044}_{-0.050}[0.851]$ & 0.21 \\[1ex]
\hline \vspace{-2ex} \\ 
\multicolumn{7}{l}{\textbf{\texttt{AbacusSummit} DESI-like $dN/dz$ \& full-sky -- DESI $0.8 < z < 2.1$ errors}}\\[1ex]
\texttt{CLEFT}& 
$0.358^{+0.041}_{-0.053}[0.346]$ & 0.67& $0.825^{+0.049}_{-0.049}[0.827]$ & 0.28 & $0.867^{+0.051}_{-0.059}[0.868]$ & 0.44 \\[1ex]
$ +\,  \Omega_m^{\rm DR1}$ prior & 
$0.318^{+0.013}_{-0.013}[0.317]$ & 0.30& $0.844^{+0.044}_{-0.044}[0.825]$ & 0.15 & $0.868^{+0.045}_{-0.045}[0.844]$ & 0.15 \\[1ex]
\hline
\end{tabular}
\caption{Parameter recovery tests on Poisson-sampled Gaussian realizations and downsampled \Abacus mocks from the angular clustering of quasars and the and quasar--CMB lensing cross-correlation. The Gaussian mocks use as fiducial input $\Omega_m=0.3164$, $\sigma_8=0.8285$, and $S_8=0.8509$. The best-fit values are the MAPs (in squared brackets) obtained from the minimizer run on top of a chain.  The values for the \texttt{AbacusSummit} downsampled simulations are the means of 18 realizations with an input cosmology centered at $\Omega_m=0.3152$, $\sigma_8=0.8080$, and $S_8=0.8282$. We quantify the modeling biases on \Abacus simulations using the pre-unblinding data error and the MAP values. Since we use parameter-level blinding, these errors are the same as the baseline errorbars from Tab.~\ref{tab:lmax_allzs} (when including BAO data) and from Eq.~\eqref{eq:noBAO_baseline} (without BAO data).}
\label{tab:simulation_results}
\end{table}

We  validate our likelihood pipeline by performing two tests: we use as input to the likelihood (i) the fiducial, noiseless input spectra with error bars taken from the DESI $0.8 < z < 2.1$ data; (ii) measured spectra from the Gaussian mocks with realistic noise levels given by $N_{\ell}^{\kappa \kappa}$ and application of the DESI survey geometry. The results of the  cosmological parameter recovery tests are summarized in table~\ref{tab:simulation_results}. For the Gaussian theory vector we find unbiased results at the $\sim 0.1\sigma$ level. Using $N=100$ Gaussian realizations of the theory input power spectrum, and applying the DESI survey geometry, we measure the resulting cosmological parameters (second row in table~\ref{tab:simulation_results}). Using the fiducial variance, we find a $\sim 0.3 \sigma$ level agreement from the mean of 100 inferred sets of parameters. This serves as a test of both our likelihood pipeline (we use an independent pipeline to generate the input theory vectors) and to quantify the impact of prior effects. 

We test our likelihood pipeline for both projection effects: first, prior ``volume effects'' where the posterior mean is found to deviate from the MAP value due to the influence of a number of prior-dominated parameters, usually occurring for weakly constraining data, and, second, prior ``weight effects'' when the true value of a parameter is ruled out (or highly disfavored) by the prior. We do not find evidence for both effects here by finding (i) agreement between the mean and the maximum of the posterior (MAP) of each chain. The MAPs are given in square brackets jointly with the means when we quote parameter results. (ii) We find negligible differences in MAP values when using uninformative, flat or wide Gaussian priors. 

\subsection{Parameter recovery tests on \Abacus mocks} \label{sec:Abacus_simulations}
The second series of tests is performed on a large suite of high-resolution gravity-only N-body \textsc{AbacusSummit} simulations\footnote{Publicly available at \url{https://abacussummit.readthedocs.io/en/latest/abacussummit.html}.} \cite{10.1093/mnras/stab2482, 10.1093/mnras/stab2484}. 
We use the Abacus lightcones \cite{HadzhiyskaLightcones} with Born-approximation CMB lensing mocks \cite{HadzhiyskaLensing}, and populate the quasars into one of the ``huge'' boxes to allow full-sky coverage out to $z = 2.1$ without any box repetitions.
This box has $8640^3 = 645$ billion particles in a $(7.5\hinvGpc)^3$ volume with an approximate particle mass of $5.7 \cdot 10^{10}  M_\odot/h$. This realization gives the density field and dark matter halos\footnote{Halos are identified using the \textsc{CompaSO} halo finder (see \cite{2022MNRAS.509..501H} for more information).} in a cubic box and lightcone using a \Planck 2018 best-fit $\Lambda$CDM cosmology \cite{Planck2018}. The halos are populated with galaxies using a halo occupation distribution (HOD) model based on the \textsc{AbacusHOD} algorithm \cite{10.1093/mnras/stab3355}, and the quasar HOD matches that from fits to small-scale clustering from the DESI early data release \cite{Yuan:2023ezi}. Since quasars are episodic, their number density is much lower than the number density of their host halos (i.e.\ they are much rarer than their clustering with linear bias of $\sim$2 would indicate) \cite{MartiniWeinberg01}. Matching the quasar number density is accomplished by randomly downsampling the parent halos. 

We therefore create two versions of the mocks; the first, the ``no downsampling'' mocks, omits this step to obtain mock quasars with a much higher number density, but a different $dN/dz$ from the DESI sample. The second set of mocks are randomly downsampled to match the DESI quasar number density as a function of redshift. To obtain mocks with as low noise as possible, we use a single realization of the ``no downsampling'' mocks on the full sky with no CMB lensing noise, and for the ``downsampling'' mocks, we average over 18 realizations of the downsampling from the single box. The no downsampling mocks have a statistical power 20 times larger than the data (as measured by the signal-to-noise of the cross-correlation at $\ell < 600$), mainly due to the lack of CMB lensing noise.
When fitting both mocks, we use the same covariance as the $0.8 < z < 2.1$ data.
A drawback of the employed simulation is that we can only test the $0.8 < z < 2.1$ bin due to the size of the simulation box; however, this redshift bin dominates our constraints.

After validating the window function treatment on the Gaussian realizations of an input power spectrum with the DESI survey applied to it, we perform a series of parameter recovery tests on \Abacus mocks. Therefore, we compare two different cases both using error bars from the low-$z$ DESI bin and the full sky with (i) no-downsampling; and (ii) DESI-like $dN/dz$. For each we run our baseline model and include an $\Omega_m$ prior centered at the \Abacus input cosmology with the width from the DESI BAO results, $\sigma_{\Omega_\textrm{m}} = 0.015$ \cite{DESI:2024mwx}. The input cosmology is $A_s = 2.083 \times 10^{-9}$, \( \Omega_m = 0.316 \), \( \sigma_8 = 0.808 \), \( S_8 = 0.828 \), \( h = 0.6736 \). The results are given in Table~\ref{tab:simulation_results}. The addition of the $\Omega_m$ prior breaks the degeneracy between $\sigma_8$ and $\Omega_m$ and (slightly) reduces biases in the pipeline to deviations of $\simlt 0.35\sigma$. We note that given that the \Abacus simulations are full-sky, the constraints are tighter than what we get from the Gaussian mocks or the DESI data. We quantify the modeling biases on \Abacus simulations using the pre-unblinding data error. Since we use parameter-level blinding, these errors are the same as the baseline errorbars from Tab.~\ref{tab:lmax_allzs} (when including BAO) and from  from Eq.~\eqref{eq:noBAO_baseline} (without BAO).

\subsection{Systematic contaminants} 
\label{sec:systematics}
\begin{figure}
    \centering
    \includegraphics[width=0.49\linewidth]{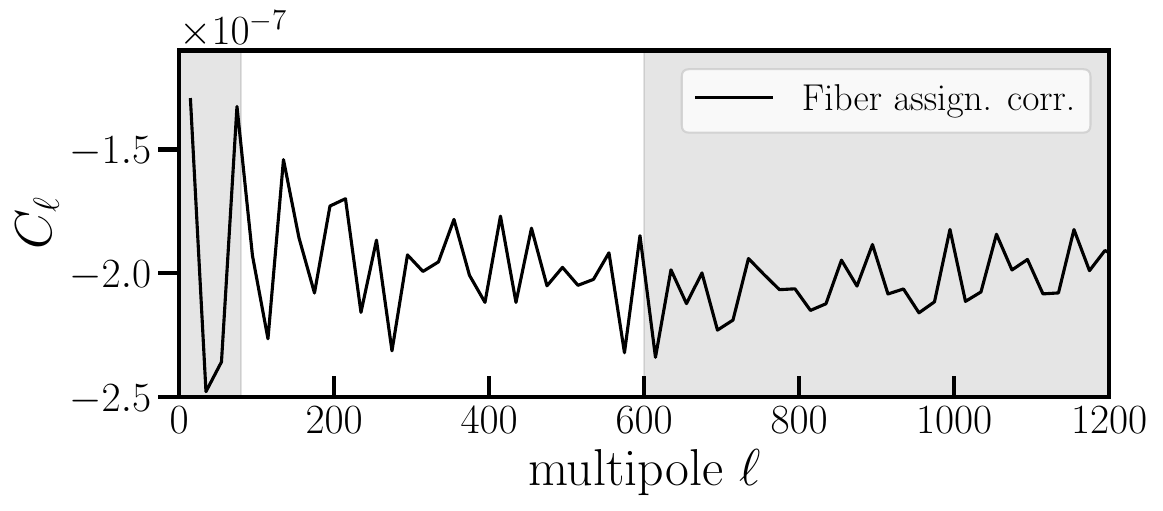}\hfill 
    \includegraphics[width=0.49\linewidth]{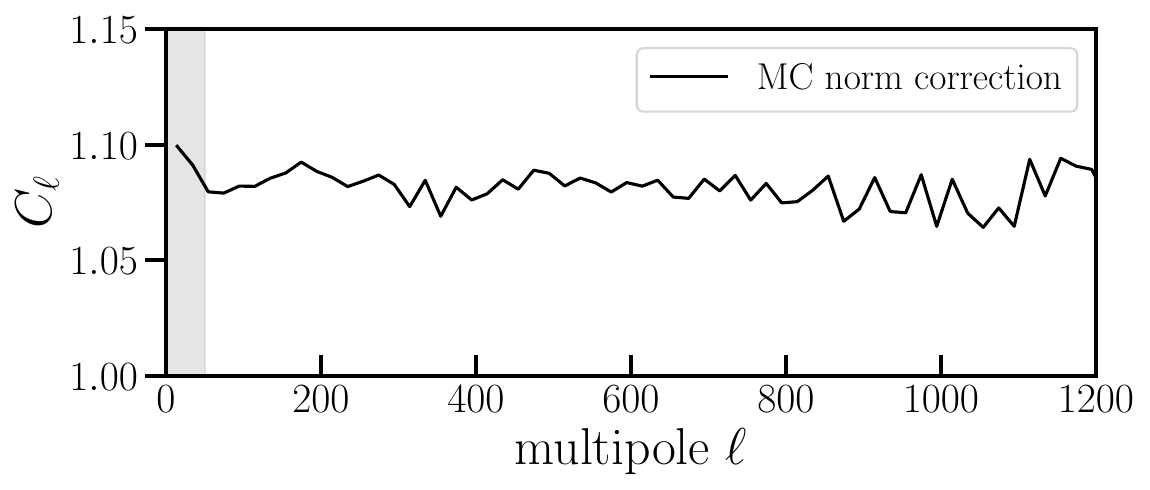}
    \caption{Transfer functions for systematic corrections applied to the data. \textit{Left panel:} Corrections from the fiber assignment applied to the auto spectrum $C_\ell^{g_1g_1}$ of the lowest redshift bin. The gray shaded regions illustrate the scale cuts applied to the measurement. \textit{Right panel:} Monte Carlo norm correction applied to the cross-spectra $C_\ell^{\kappa g_i}$ which amounts to an approximately constant amplitude change of $7-10 \%$.}
    \label{fig:systematic_corrections}
\end{figure}

In this section we present and discuss a series of tests performed to demonstrate robustness against systematic contaminants. 
\begin{enumerate}
    \item \textbf{Fiber assignment corrections}: We estimate the impact of fiber assignment by computing the angular power spectrum of the \Abacus mocks used to validate the DESI DR1 RSD modelling \cite{DESI:2024hhd}.
    These are distinct mocks from the \Abacus huge-box lightcones used to validate the angular clustering measurements. They are constructed from the 25 \Abacus ``base'' boxes of size $(2\hinvGpc)^3$ with remapping to fit the entire DESI volume, and are described in Section 3.1 of \cite{DESI:2024hhd}. These mocks are available in two flavors, ``complete'' in which the DESI footprint is applied but fiber-assignment is not run, and ``Alt-MTL'' \cite{Lasker25} in which the DESI \texttt{FIBERASSIGN} code was run on the mocks. We measure the average angular power spectra for $0.8 < z < 2.1$ quasars in  both flavors and plot its difference $\Delta C_\ell^{\textrm{AltMTL}}$, in the left panel of Fig~\ref{fig:systematic_corrections}. We find that, on the scales of interest, $\Delta C_\ell^{\textrm{AltMTL}}$ is negative (i.e.\ fiber assignment reduces clustering power as it prevents observation of small-scale galaxy pairs), and consistent
   with a constant offset at a level of $\sim$10\% of the shot-noise level.\footnote{$\Delta C_\ell^{\textrm{AltMTL}}$ becomes scale-dependent when considering much smaller scales, $\ell > 2000$.} As a result, we expect that our shot-noise marginalization (with width equal to a third of the fiducial shot-noise value) will capture the impact of fiber collisions on the angular clustering.
    We explicitly test this by subtracting $\Delta C_\ell^{\textrm{AltMTL}}$
    from the data and re-running our likelihood we find a $-0.05\sigma$ change in \sig (and equivalently in \Sig).
    Because of the minimal impact, we do not wish to add this noisy correction to the data, and therefore do not apply any correction in our baseline analysis.
    We do not include a correction in the cross-spectra, following measurements of galaxy-galaxy lensing on fiber-assigned mocks shown in Fig.~7 in Ref.~\cite{Lange:2024vmv}, who find that applying completeness weights is adequate to eliminate the impact of fiber collisions on the cross-correlation.
    \item \textbf{Monte-Carlo norm correction to \Planck lensing maps:} Following \cite{Sailer2024, ACT:2023oei}, we apply a simulation-based mode-by-mode correction to the measured cross-spectrum. This is required as the mode-coupling stemming from anisotropic filtering and masking (both of which are applied during lensing reconstruction) is not forward modeled into the cross-spectra obtained with \textsc{NaMaster} algorithm. As this correction factor is mask-specific, we generate a suite of CMB lensing reconstructions with the appropriate mask
\begin{equation}
    T_{L} \equiv \frac{\sum_{\ell} W_{L\ell}\sum_{i}^{N_{\rm sim}} \sum_{m=-\ell}^{\ell}\{M^{\kappa}\kappa^{i}\}_{\ell m} \{M^{q}\kappa^{i}\}_{\ell m}^{\ast}}{\sum_{\ell} W_{L\ell}\sum_{i}^{N_{\rm sim}} \sum_{m=-\ell}^{\ell}\{M^{\kappa}\Hat{\kappa}^{i}\}_{\ell m} \{M^{q}\kappa^{i}\}_{\ell m}^{\ast}}\, ,
\end{equation}
where $M$ are the masks, $\kappa$ the input ($\Hat{\kappa}$ the reconstructed) lensing convergence, $i$ is an index over simulations, and $\{XY\}_{\ell m} \equiv \int \diff^2 \Hat{\mathbf{n}} Y_{\ell m}^{\ast}(\Hat{\mathbf{n}}) X(\Hat{\mathbf{n}})Y(\Hat{\mathbf{n}})$ with $Y_{\ell m}$ being the spherical harmonics. The corrections are $7-10\%$ for the \Planck PR4 maps over the DESI DR1 footprint at the scales of interest, see right panel in figure \ref{fig:systematic_corrections}. We do not find evidence for a scale dependent MC norm correction. 
    \item \textbf{Catastrophic redshift failures:} Accurately measuring redshifts for quasars is challenging. Whilst this is a key concern for 3D analyses \cite{Chudaykin:2022nru, Chaussidon:2024qni}, the large width of the here chosen redshift bins reduces its impact. 
    \item \textbf{Angular integral constraint:} Methods to correct for imaging systematics can spuriously remove power on large angular scales. As described in Section 10.1.2 of \cite{DESI2024II}, this effect is estimated by applying the imaging systematics correction to cutsky mocks without imaging systematics contamination and measuring the change in the power spectrum monopole and quadrupole. This change in power is termed the ``angular integral constraint'' (AIC). For 3D galaxy clustering, \cite{DESI2024II} finds that linear regression removes negligible large-scale power, while the nonlinear methods used (SYSNET for ELGs and Random Forest for QSOs) require an AIC correction.
    In this work, we use linear weights rather than the RF weights to mitigate the impact of the AIC.
    Additionally, we check the AIC by measuring the angular power spectrum of 40 EZMocks with a $6h^{-1}$ Gpc volume  (as used in \cite{Chaussidon:2024qni}) both with and without the linear regression weights. We find a $<0.2\%$ change in the auto-spectrum on the scales used ($\ell > 85$), although at $\ell \sim 15$ linear regression reduces $C_\ell^{gg}$ by $\sim$2\%. We expect that the impact of the AIC on the cross-correlation is less than the impact on the auto-correlation, which is $\sim$0.5\% at $\ell \sim 40$.
    \item \textbf{Quasar cross-correlation spectra:} 
    Whilst $g_i$ and $g_j$ are, in principle, uncorrelated, magnification bias creates a non-zero correlation. 
    We find that the cross-spectra agree with the theory prediction in our fiducial scale range ($\ell > 85$), once a fiber-assignment correction of $\Delta C_\ell^{\textrm{AltMTL}} \sim 2 \times 10^{-7}$ is added.
    On larger scales, we find reasonable agreement at $\ell \geq 35$ for the $g_1$-$g_2$ cross-correlation, but disagreement at $\ell < 85$ for $g_1$-$g_3$ and $g_2$-$g_3$. This may be due to residual imaging systematics or large-scale discrepancies due to our pixelized power spectrum estimator, and the large-scale inter-bin quasar cross-correlation would be interesting to explore with the pixel-free approach as well.
    \end{enumerate}
    
\section{Results} \label{sec:results}
In this section we present a tomographic measurement of the amplitude of matter fluctuations over the redshift range $0.8 \leq z \leq 3.5$ in three redshift bins from the cross correlation of 1,223,391 spectroscopically confirmed quasars selected by the Dark Energy Spectroscopic Instrument (DESI) across 7,200 deg$^2$ and Planck PR4 (NPIPE) cosmic microwave background (CMB) lensing maps. We use the angular power spectrum estimator, described in section \ref{sec:NaMaster}, to measure the auto and cross-spectra and use the Convolutional Lagrangian Effective Field Theory (CLEFT), described in section \ref{sec:theory} as our theoretical model. In section \ref{sec:results} and figure \ref{fig:sigma8_z} we show the key results of this paper: tomographic constraints on the amplitude of matter clustering centered at mean redshifts $\overline{z} \sim 1.5,\, \text{and}\, 2.9$.  For the present analysis we quote the results for the joint analysis of the northern and southern galactic cap (NGC/SGC), weighted by the number of quasars in each hemisphere.

\subsection{Blinding} \label{sec:blinding}
We performed a \textit{blind} analysis on the cosmological constraints derived from the cross-correlation of DESI DR1 quasar data with \Planck PR4. First, we visually inspected the measured angular auto- and cross-spectra (but did not overplot theory and data vectors). Second, we tested and validated the scale cuts and priors on Gaussian and \Abacus mocks. Once these were passed, these values were frozen and we performed a blind cosmological analysis of the present data (blinded at the likelihood level). After the pre-defined blinding tests were satisfactorily passed, we unblinded the cosmological constraints. We emphasize our baseline analysis choices have \textit{not} changed post unblinding, in particular, the scale cut $\ell_{\rm max}=605,\,1205,\,1205$ compared to the more conservative scale cut of $\ell_{\rm max}=605,\,605,\,605$. The only measurement we did not validate on simulations was the sound-horizon free measurement of $H_0$ since we do not have simulations at hand for this.  To ensure robust results we tested the chosen analysis variations (different priors and scale cuts) additionally post unblinding and did not find any reasons for concern in the choices of our scale cuts. 

\subsection{Best-fit angular power spectra}
\begin{figure}
    \centering
    \includegraphics[width=1\linewidth]{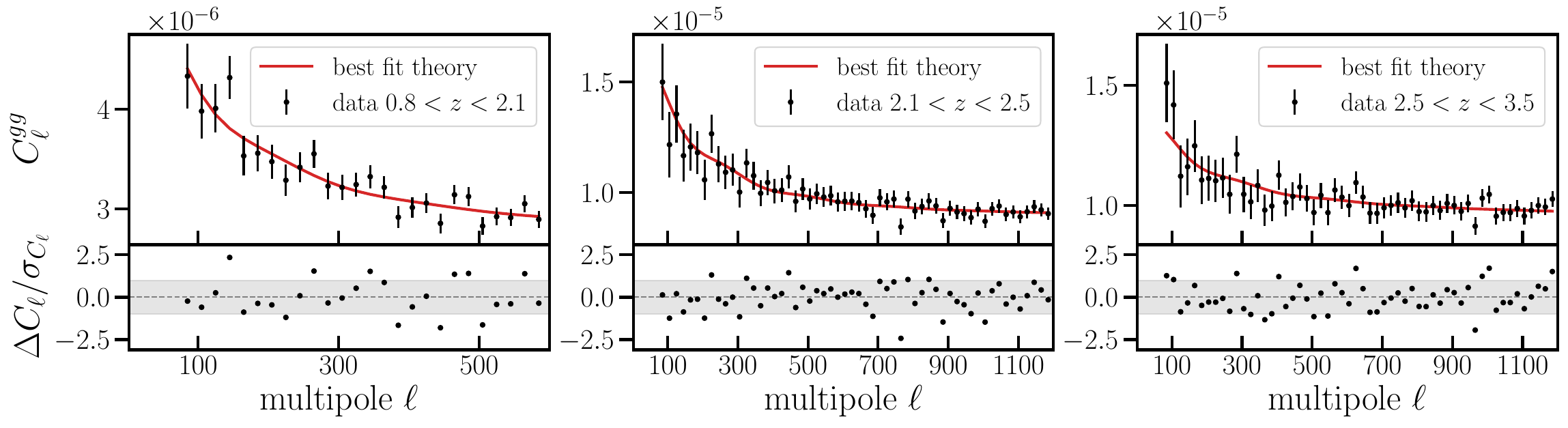}\\
    \includegraphics[width=1\linewidth]{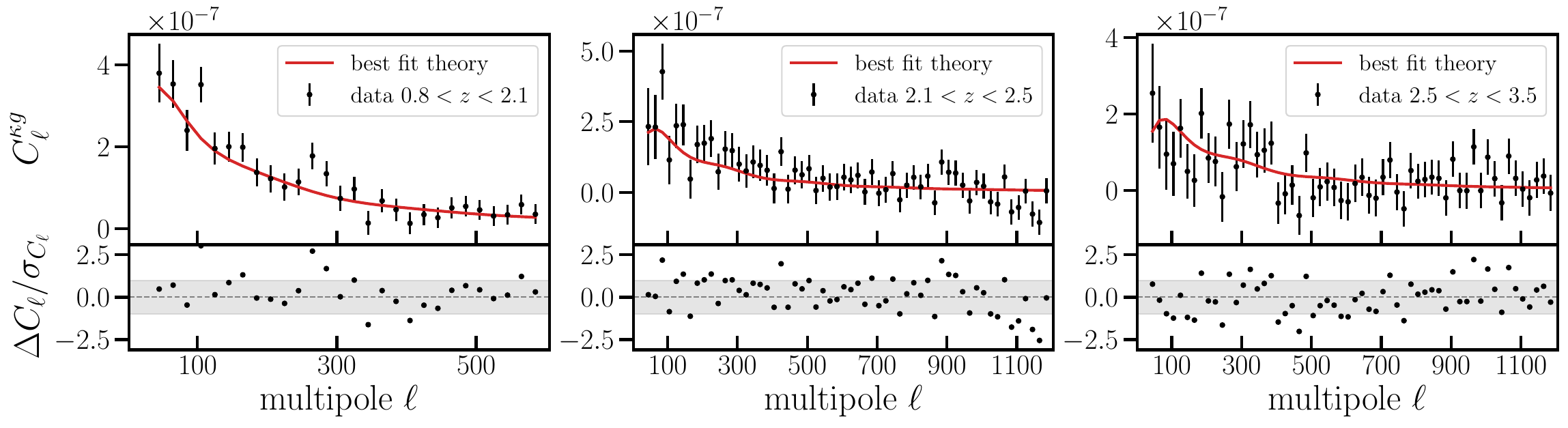}
    \caption{Comparison of CLEFT best-fit angular power spectra (red line) obtained from the joint analysis of the three redshift bins to the measurement points (black dots) including DESI DR1 BAO data. The redshift is increasing from left to right with scale cuts of $\ell_{\rm max}=605,\, 1205,\, 1205$ for each redshift bin. The error bars are from the square root of the diagonal of the covariance. \textit{Top panel:} We show the auto-spectra with scale cut of $\ell_{\rm min}=85$. The $\sigma$-normalized residuals are shown in the bottom part of each plot with a gray $1\sigma$ band to guide the eye. \textit{Bottom panel:} We show the cross-spectra with scale cut of $\ell_{\rm min}=45$. The goodness of fit is $\chi^2=273.7$ for 282 data points and 14 free parameters, tabulated in Tab.~\ref{tab:results}.}
    \label{fig:bestfit_Cell}
\end{figure}

In Fig.~\ref{fig:bestfit_Cell} we show the best fit angular power auto (top row) and cross-spectra (bottom row) from a joint fit of the three redshift bins. In each plot we show the residuals normalized by their error bars with a gray $1\sigma$ error band to guide the eye. We find good agreement between the model and the data at the $1-1.5\sigma$ level with a $\chi^2$ value of 273.7 for 282 data points and 14 degrees of freedom.\footnote{Note analytically marginalizing over the shot noise requires a careful treatment of the $\chi^2$ statistic (see, e.g.,~\cite{2024arXiv240704607S} for more details).} For the auto-(cross-)spectra we apply a scale cut of $\ell_{\rm min}=85\, (45)$. Variations of the scale and redshift bins are given in table \ref{tab:lmax_allzs} and, following baseline expectation, show good consistency. We do not see a strong improvement when increasing $\ell_{\rm max}$, which indicates that our results are shot noise dominated for $\ell>800$. 

For our fiducial analysis, we detect the cross-correlation between PR4 lensing maps and DESI DR1 with a signal-to-noise ratio of $27.2$ for the auto-correlation and $21.7$ for the cross-correlation for the joint analysis of all redshift bins. This corresponds to SNRs of $20.0,\, 15.8,\, 9.6$ ($19.8,\, 8.7,\, 4.9$) for the individual bins in ascending order for the auto- (cross-) correlation, respectively.  

\subsection{Cosmological constraints from tomographic analysis}

\begin{table}
\centering
\setlength{\tabcolsep}{4pt}
\begin{tabular}{cccccccccc}
\hline \hline \vspace{-2ex} \\ 
$\ell_{\mathrm{max}}^{z_1}$ &$\ell_{\mathrm{max}}^{z_2}$ &$\ell_{\mathrm{max}}^{z_3}$ & $A_\mathrm{s} \cdot 10^{9}$ & BAO & $\Omega_m$ & $\sigma_8$ & $S_8$ &$\chi^2$ & $N_d$ \\[1ex]
\hline \vspace{-2ex} \\ 
605 & -- & -- & $2.73^{+0.34}_{-0.52}$ & DR1 &$0.296 \pm 0.007$ & $0.910^{+0.063}_{-0.083}$ & $0.904^{+0.064}_{-0.082}$ & 42.5 &54 \\[1ex]
605 & -- & -- & $2.69^{+0.33}_{-0.53}$ & DR2 &$0.297 \pm 0.004$ & $0.905^{+0.062}_{-0.086}$ & $0.901^{+0.061}_{-0.085}$ & 43.3 &54 \\[1ex]
\hline \vspace{-2ex} \\ 
605 & 605 & 605 & $2.95^{+0.34}_{-0.57}$ & DR1 &$0.296 \pm 0.006$ & $0.946^{+0.062}_{-0.088}$ & $0.938^{+0.061}_{-0.088}$ & 148.9&162 \\[1ex]
605 & 805 & 805 & $2.88^{+0.33}_{-0.52}$ & DR1 &$0.295 \pm 0.006$ & $0.935^{+0.060}_{-0.081}$ & $0.928^{+0.060}_{-0.081}$ & 187.8& 202 \\[1ex]
605 & 1205 & 1205&$2.84^{+0.33}_{-0.48}$ & DR1 &$0.296 \pm 0.006$ & $0.929^{+0.059}_{-0.074}$ & $0.922^{+0.059}_{-0.073}$ & 273.7 & 282 \\[1ex]
\hline \vspace{-2ex} \\ 
605 & 605 & 605 & $2.89^{+0.36}_{-0.52}$ & DR2 &$0.297 \pm 0.004$ & $0.938^{+0.065}_{-0.082}$ & $0.933^{+0.064}_{-0.081}$ & 149.2&162 \\[1ex]
605 & 805 & 805 & $2.91^{+0.33}_{-0.54}$ & DR2 &$0.297 \pm 0.004$ & $0.941^{+0.059}_{-0.085}$ & $0.936^{+0.058}_{-0.085}$ & 186.4& 202 \\[1ex]
605 & 1205 & 1205&$2.86^{+0.34}_{-0.53}$ & DR2 &$0.297 \pm 0.004$ & $0.934^{+0.062}_{-0.082}$ & $0.930^{+0.061}_{-0.082}$ & 273.5 & 282 \\[1ex]
\hline
\end{tabular}
\caption{Scale and prior dependence of the cosmology constraints as a function of $\ell_{\rm max}$ for the lowest redshift bin and the joint analysis of all $z$-bins. These results use the baseline \textsc{CLEFT} model with BAO information from DESI DR1 or DR2, $\Omega_mh^3 = 0.09633$ and $n_s=0.9649$ fixed to the \Planck best-fit value \cite{Planck:2018vyg} and $\omega_b$ to the Big Bang Nucleosynthesis value \cite{Schoneberg:2019wmt}. The inclusion of the BAO data from DESI DR1/DR2 breaks the \sig-$\Omega_m$ degeneracy but results in prior dominated results on the matter density. The corresponding best-fit spectra are shown in figure~\ref{fig:bestfit_Cell}. In the last two columns we quote the goodness of fit and the corresponding number of data points. Each fit has 14 free parameters. Although we substantially increase the $\ell_{\rm max}$, the information gain is minimal without a deterioration of the $\chi^2$.} \label{tab:lmax_allzs}
\end{table}

\begin{figure}
    \centering
    \includegraphics[width=\linewidth]{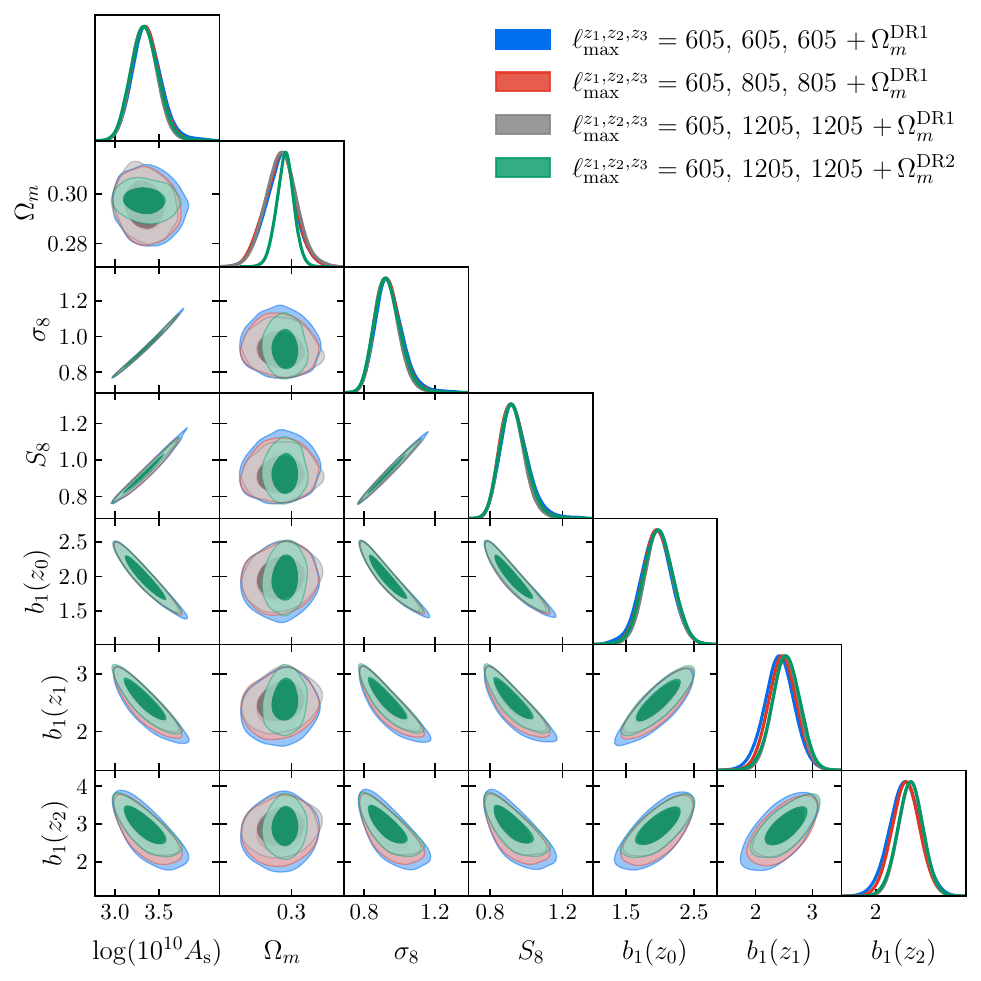}
    \caption{Triangle plot of the cosmology constraints obtained from cross-correlation of DESI DR1 quasars in three redshift bins from $0.8\leq z \leq 3.5$ using a CLEFT theory model as a function of scale cut $\ell_{\rm max}$. We impose a prior on $\Omega_m$ from the DESI DR1 (blue, red, gray contours) and DR2 (green contours) BAO measurement \cite{DESI:2024mwx, DESI:2025zgx} to break the $\Omega_m - \sigma_8$ degeneracy.}
    \label{fig:DESI_triangle}
\end{figure}

In figure ~\ref{fig:DESI_triangle} we show a triangle plot of the marginalized 1D and 2D posteriors for the cosmological parameters as a function of scale cut. The corresponding numerical values are given in table~\ref{tab:lmax_allzs}. We do not quote the results for the two higher redshift bins as the low SNRs yield very poor cosmological constraints which are improved  when performing fits to the joint $z$-bins. Further, for the lowest redshift bin we only quote results up to $\ell_{\rm max}=805$ as we do not trust our theory modeling up to these small scales at these redshifts. We find evidence for prior volume effects (also from Gaussian and \Abacus simulations) as the MAP values are in $0.3-0.8\sigma$ tension with the means of the chains.\footnote{We present an analysis variation using linear theory in Sec.~\ref{sec:lin_theory} to address this.} Our baseline results for the highest $\ell_{\rm max}$ constraint (gray contours) are given in the last row of Table \ref{tab:lmax_allzs}. 
We find a small gain increasing the scale cuts from $\ell_{\rm max}=605$ to 1,205 which is consistent with the fact that we are shot noise dominated at these scales. This stems from the sparse sampling of the quasars at high redshifts. Following baseline expectation, the constraints on the matter density and $S_8$ degrade when \textit{not including} BAO to 
\begin{subequations} \label{eq:noBAO_baseline}
\begin{align}
    &A_s =\left(2.73^{+0.36}_{-0.55}\, [2.93]\right)\cdot 10^{-9},\, 
    &\Omega_m = 0.308\pm 0.048\, [0.301]\, ,\\ 
    &\sigma_8 = 0.913^{+0.058}_{-0.075}\, [0.986],\, 
    &S_8 =0.920^{+0.076}_{-0.086}\, [0.988]\, .
\end{align}
\end{subequations}

We include the second order bias parameters in the CLEFT framework, namely $b_2$ and $b_s$. Whilst these are only weakly detected these are dictated by the symmetry of the tracer and include a marginalization over all possible bias parameter values. For our baseline results, we measure the bias parameters per redshift bin. For the linear bias parameter we find: 
\begin{equation} \label{eq:bias_results}
    b_1(z_1) = 1.98 \pm 0.20\,, 
    b_1(z_2) = 2.53 \pm 0.23\,, 
    b_1(z_3) = 2.93 \pm 0.36\,.
\end{equation}
The constraints from this work are discrepant with \cite{Chaussidon:2024qni} at the $1.3,\, 3.6,\,  3.8 \sigma$ level. The higher order bias parameters yield 
\begin{equation}
    b_2(z_1) = (\mathrm{prior})\,, 
    b_2(z_2) = 0.2^{+1.6}_{-1.1}\,,
    b_2(z_3) = 0.06^{+0.94}_{-1.1}\,,
\end{equation}
where for $b_2(z_1)$ we recover the flat prior similarly for the shear bias for which we recover the prior
\begin{equation}
    b_s(z_1) = -0.10 \pm 1.00\,, 
    b_s(z_2) = 0.00 \pm 1.00\,, 
    b_s(z_3) = 0.00 \pm 1.00\, .
\end{equation}
For the magnification bias, we find consistent values to our measurements on the data
\begin{equation}
    s_{\mu}(z_1) = 0.10 \pm 0.03\,, 
    s_{\mu}(z_2) = 0.22 \pm 0.05\,, 
    s_{\mu}(z_3) = 0.19 \pm 0.06\,.
\end{equation}
For the shot noise, which we can recover \textit{a posteriori} from the chain, we find lower values than measured on the data 
\begin{equation}
    \mathrm{SN}(z_1) = (2.73 \pm 0.02)\cdot 10^{-6}\,, 
    \mathrm{SN}(z_2) = (8.77 \pm 0.05)\cdot 10^{-6}\,, 
    \mathrm{SN}(z_3) = (9.58 \pm 0.05)\cdot 10^{-6}\,. 
\end{equation}

\begin{figure}
    \centering
    \includegraphics[width=1\linewidth]{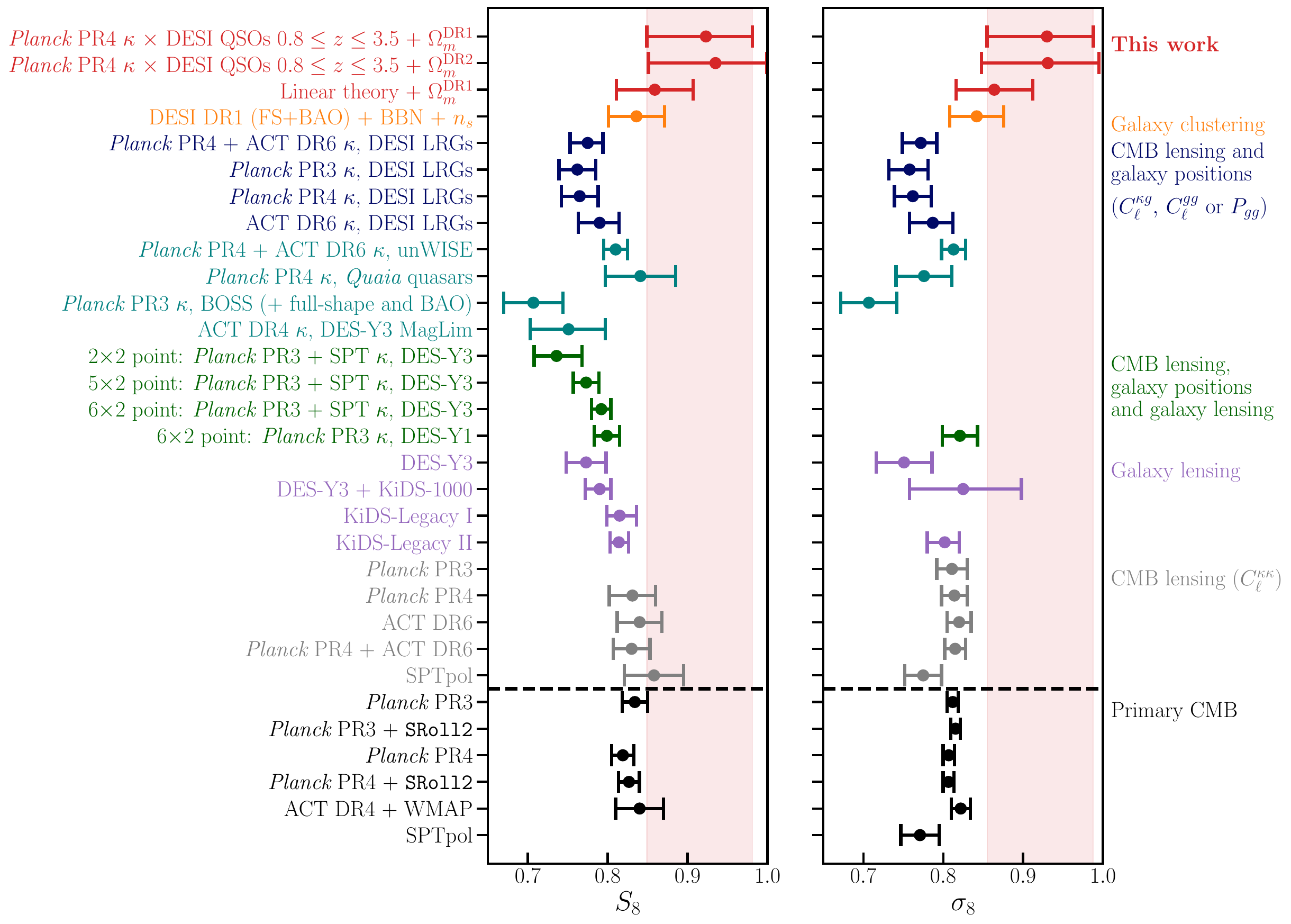}
    \caption{Comparison of constraints on the growth of structure analysis variations presented in this work (first three rows in red) to other lensing cross-correlation analyses. We show constraints on the growth of structure from the cross-correlation of CMB lensing and (i) galaxy positions (dark blue and teal), (ii) 3D galaxy positions and lensing (dark green), (iii) galaxy lensing (purple), (iv) CMB lensing and (v) from the primary CMB (black). For $\sigma_8$ and $S_8$, our baseline (linear theory; see Sec.~\ref{sec:lin_theory}) results are $\sim 1.6\sigma$ ($\sim 1\sigma$) higher than predictions from $\Lambda$CDM fits to measurements of the primary CMB from \Planck PR4.}
    \label{fig:compilation_S8}
\end{figure}

In Fig.~\ref{fig:compilation_S8} we show a compilation of different growth of structure measurements using $\sigma_8$ and $S_8\equiv \sigma_8(\Omega_m/0.3)^{0.5}$. Our results are consistent with recent measurements at the $1-2.5\sigma$ level\footnote{Given the asymmetric error bars, we use the left-sided error bar to compute the ``tension'' $T$  to quantify in units of $\sigma$ the discrepancy between two measurements: $T=(\mu_1-\mu_2)/\sqrt{\sigma_{\rm low}^2}$. 
Whilst this metric is overly simplistic as it assumes a  Gaussian posterior and ignores correlations between datasets it is a helpful tool to approximately assess the level of consistency between different measurements. We verified that using a more sophisticated technique accounting for the non-Gaussianities in the parameter posteriors \cite{Raveri:2021wfz} does not increase the tension with \Planck 2018 ($\texttt{TT,TE,EE+lowE}$).} with measurements from the primary CMB: our joint constraints are $\sim 1.5 \sigma$ high compared to \Planck 2018 ($\texttt{TT,TE,EE+lowE}$) \cite{Planck:2018vyg} even when combined with a low-$\ell$ \texttt{SRoll2} likelihood (\texttt{momento} \cite{deBelsunce:2022yll}) which includes cross-correlations between temperature and polarization in de Belsunce \emph{et al.} \cite{deBelsunce:2021mec}; similarly for the most recent PR4 analysis in Tristram \emph{et al.} \cite{Tristram:2023haj}. \Sig measurements are discrepant at the $1.0-1.4\sigma$ level from $\Lambda$CDM predictions obtained from measurements of the primary CMB. This discrepancy is reduced by $\sim 0.5\sigma$ when using our linear theory analysis variation (see Sec.~\ref{sec:lin_theory}). 

The most relevant measurements to compare our results to are (i) Sailer \emph{et al.} \cite{2024arXiv240704607S, 2024arXiv240704606K} which present cross-correlation of DESI LRGs and \Planck PR4 + ACT DR6 measurements; (ii) Farren \emph{et al.} from cross-correlation of the same lensing maps with unWISE blue and green samples \cite{ACT:2023oei,Farren:2024rla}. In particular, their hybrid EFT model is very similar to our theory model. (iii) Piccirilli \emph{et al.} \cite{Piccirilli:2024xgo} cross-correlating PR4 and \textit{Quaia} quasars \cite{Storey-Fisher:2023gca}. (iv) Miyatake \emph{et. al.} correlating Lyman Break Galaxies with PR3 lensing maps \cite{Miyatake:2021qjr}. Using individual redshift bins, our $z_3$ measurement presents one of the highest redshift measurement of \sig (\Sig) to date. For the joint constraints, we are centered at a similar redshift to (ii) and (iii). Our results are $\sim 2 \sigma$ higher than each of the measurements, respectively. We do not find any conclusive evidence for a \sig (\Sig) tension with the DESI DR1 quasar sample centered at $z\approx 2.3$. Note that our constraints on $\sigma_8$ are fully consistent with the results obtained from the ``full modeling'' given in table 10 from the DESI DR1 QSO power spectrum \cite{DESI:2024jis}. 

Whilst our error bars are broader compared to Farren \textit{et al.} and Piccirilli \textit{et al.}, we find that our number density is lower than from their measurement with a wider redshift coverage. In particular, we do not use very informative priors (\ie we do not fix the counterterms to zero) but marginalize over the quadratic biases $b_2$ and $b_s$. We note that we only weakly detect them but prefer to leave them in since this matches the blinding strategy and while linear theory is powerful in this regime (as shown at lower redshift in Sailer \textit{et al.} \cite{2024arXiv240704607S}) we find deviations from linear theory on our parameter recovery tests on \textsc{AbacusSummit} simulations with the fiducial scale cuts up to $\ell_{\rm max}=605,\,1205,\,1205$. 

Our results are $1.4 - 2.2 \sigma$ high compared to measurements from galaxy lensing, as illustrated by the purple points in figure \ref{fig:compilation_S8}. We show a sub-set of (more recent) constraints from the joint analysis of Dark Energy Survey (DES-Y3)  and the Kilo-Degree Survey (KiDS-1000) \cite{KiDSDES} as well as KiDS-Legacy releases \cite{Stolzner25, Wright25}. 

\begin{figure}
    \centering
    \includegraphics[width=1\linewidth]{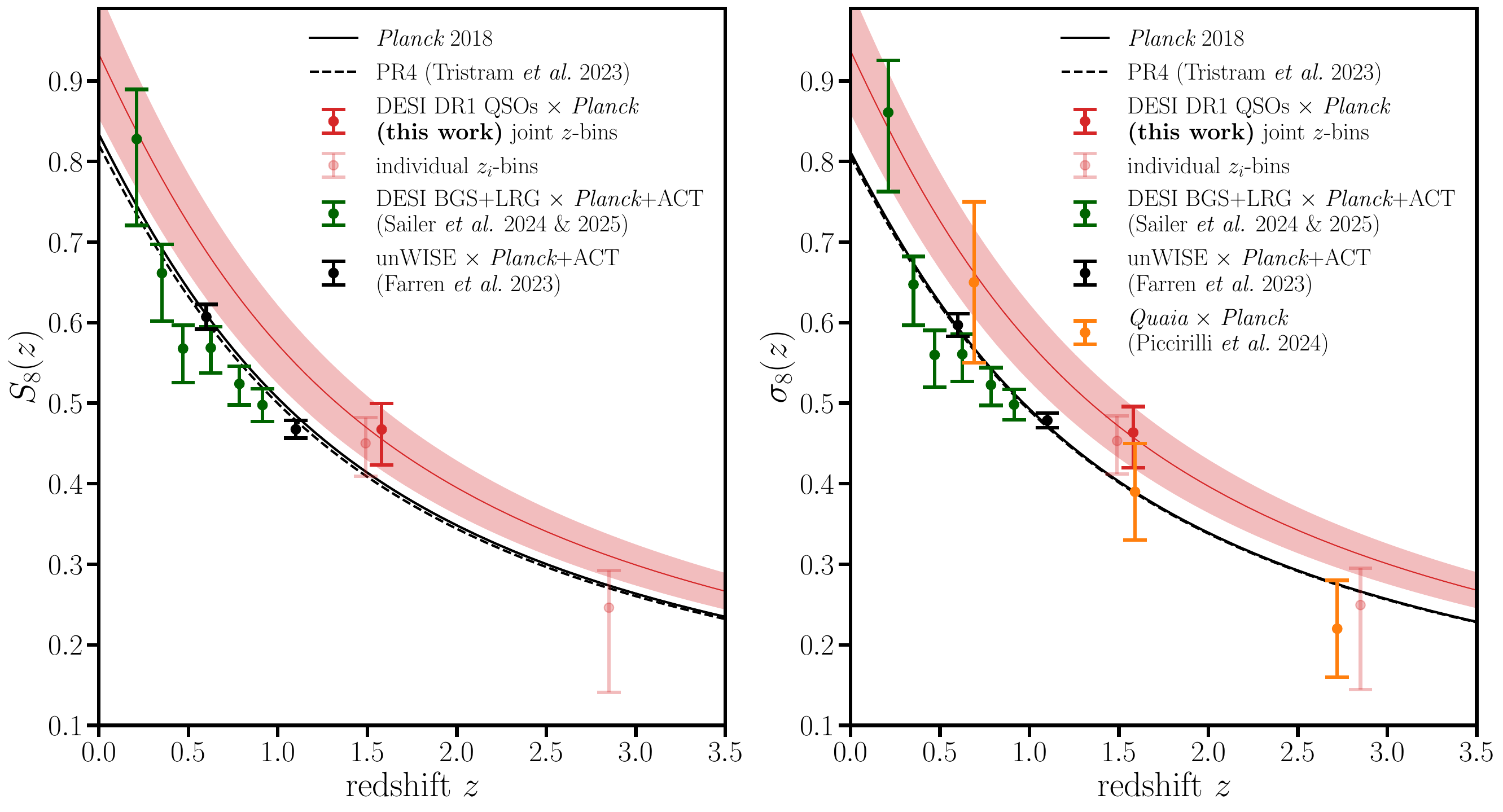}
    \caption{Comparison of selected tomographic growth of structure measurements from CMB lensing cross-correlations with galaxy and quasar positions. We show the \Planck PR4 prediction in black \cite{Tristram:2023haj}, this work is shown in red for the lowest and highest redshift bin at $z_{\rm eff}$ and for the joint analysis of all three bins. Further, we compare our results to three recent measurements: (i) \Planck PR4 + ACT DR6 cross-correlation with DESI LRG+BGS \cite{2024arXiv240704607S, Sailer:2025rks} in green; (ii) \Planck PR4 + ACT DR6 correlated with the blue and green unWISE galaxy samples \cite{ACT:2023oei} in black; and (iii) \Planck PR4 correlated with the \emph{Quaia} quasar sample in orange \cite{Piccirilli:2024xgo}. }
    \label{fig:sigma8_z}
\end{figure}

In figure \ref{fig:sigma8_z} we compare our tomographic growth of structure measurements with recent measurements from CMB lensing cross-correlation with galaxy positions.\footnote{Measurements that do not quote $\sigma_8(z)$ are rescaled by $\sigma_8(z)/\sigma_8(z=0)$ assuming a \Planck 2018 $\Lambda$CDM cosmology.} In particular, we compare the constraints that are closest to our work re theory modeling (Farren \textit{et al.} \cite{ACT:2023oei}) and chosen data sample (Piccirilli \textit{et al.} \cite{Piccirilli:2024xgo}). Our constraints are shown as red points for the first and the last redshift bin individually (the center redshift bin has poor constraining power on its own) and for the joint constraints (bold red point). Given the large sample size of low-redshift quasars, the mean redshift in our analysis approximately 1.5, almost exactly matching the high redshift bin in Farren \textit{et al.} and the center redshift bin in Piccirilli \textit{et al.}. We do not find any evidence for a departure from $\Lambda$CDM from DESI DR1 quasar cross-correlation measurements with \Planck PR4. 
Our constraining power is similar to that from 
Piccirilli \textit{et al.} at $z \sim 1.5$ and slightly worse than them at $z > 2$. The constraints from Farren \textit{et al.} are tighter, but at this redshift they also incorporate information from the CMB lensing auto-correlation, whereas our measurement and those from Piccirilli \textit{et al.}  are from cross-correlations alone.

\subsection{Analysis variation: Linear theory fits} \label{sec:lin_theory}
\begin{figure}
    \centering
    \includegraphics[width=\linewidth]{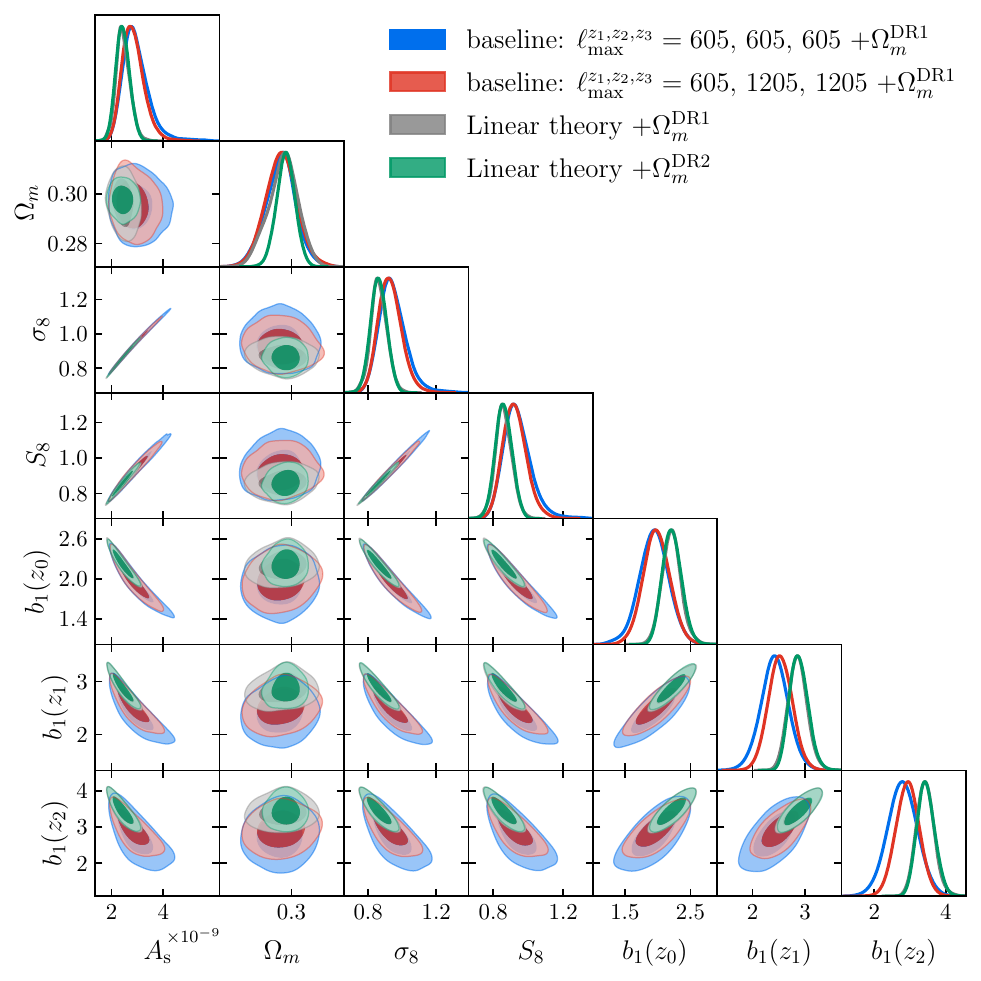}
    \caption{Triangle plot comparing cosmological constraints obtained from cross-correlation of DESI DR1 quasars in three redshift bins from $0.8\leq z \leq 3.5$ using a CLEFT theory model (baseline analysis) and linear theory fits imposing a prior on $\Omega_m$ from the DESI DR1 and DR2 BAO measurement \cite{DESI:2024mwx, DESI:2025zgx} to break the $\Omega_m - \sigma_8$ degeneracy. The linear theory fits use the same scale cuts as the blue contours. }
    \label{fig:DESI_triangle_Quaia}
\end{figure}

To explore the difference between our analysis and the one presented in Ref.~\cite{Alonso23} we fix the higher order biases, magnification bias and shot noise and only sample $b_1$ in addition to the cosmological parameters. The results are shown in Figs.~\ref{fig:compilation_S8} and \ref{fig:DESI_triangle_Quaia} with this variation denoted by ``Linear theory''. We additionally compare the impact of using the DESI DR2 BAO data \cite{DESI:2025zgx}. Following baseline expectation, we find tighter constraints when using linear theory which use as scale cut corresponding to the blue contours. Whilst our baseline analysis marginalizes over the higher-order nuisance parameters this comes at a cost of removing shape information.  For the growth of structure we find 
$\sigma_8 = 0.864\pm 0.048$ $[0.866]$ 
($\sigma_8 = 0.862\pm 0.047$ $[0.859]$) 
using a DESI DR1 (DR2) BAO prior. This corresponds to a discrepancy of $-0.99 \sigma$ ($-0.97 \sigma$). For $S_8$ we find 
$S_8 = 0.859\pm 0.048$ $[0.863]$
($S_8 = 0.859\pm 0.048$ $[0.855]$) 
a $-1.08 \sigma$ ($-1.10 \sigma$) difference. For the linear bias parameters we obtain 
$b_1(z_1) = 2.21\pm 0.15$ $[2.2]$
$b_1(z_2) = 2.86^{+0.17}_{-0.19}$ $[2.88]$ and 
$b_1(z_3) = 3.43^{+0.23}_{-0.26}$ $[3.45]$
which are in somewhat better agreement, \textit{i.e.}~$(0.21,\, 2.57,\, 3.49)\sigma$, with the values from Eq.~\eqref{eq:bias_evolution} measured from the 3D clustering of DESI DR1 quasars \cite{Chaussidon:2024qni} and the values obtained from our fiducial analysis given in Eq.~\eqref{eq:bias_results}. The DR2 constraints are within $\simlt 0.05 \sigma$. 
The constraining power of this ``linear theory'' analysis is in better agreement with that of \cite{Alonso23}, with a $\sim 5.5\%$ constraint on $S_8$ for our linear theory fits compared to a 5.1\% constraint on $S_8$ in \cite{Alonso23}. Given the similar detection significances of $C_\ell^{gg}$ and $C_\ell^{\kappa g}$ in \cite{Alonso23} (21.7$\sigma$ and 12.3$\sigma$) and our work, this agreement is encouraging. However, our fractional constraining power degrades to $\sim8\%$ on $S_8$ in our baseline configuration when we marginalize over shot noise, higher-order bias parameters, and magnification bias.

\subsection{Sound-horizon free measurement of the Hubble constant} 
\label{sec:hubble}
In light of the vigorously debated $H_0$ tension, many models have been proposed that involve the modification of the physical size of the sound horizon at recombination. Thus, measurements of the Hubble constant independent of the sound horizon have gained traction in the community (see, e.g.,~\cite{Zaborowski:2024wpo, Alonso:2024emk} and references therein). Aside from using the BAO as ``standard ruler'', we use the matter-radiation equality scale $k_{\rm eq}$ visible as the turnover scale in the matter power spectrum \cite{Philcox:2020xbv, Farren:2021grl} which is given by 
\begin{equation}
    k_{\text{eq}} = \left( 2\Omega_{\text{cb}} H_0^2 z_{\text{eq}} \right)^{1/2} \approx 7.46 \times 10^{-2} \Omega_{\text{cb}} h \Theta_{2.7}^{-2} \,,
    \left[ h \text{ Mpc}^{-1} \right]
\end{equation}
where $\Theta_{2.7} \equiv T_{\text{CMB}} / (2.7\,\text{K})$ and $\Omega_{\rm cb}$ is the sum of cold dark matter and baryon density. In the angular projection this corresponds to $\Omega_m^{0.6}/h$. We can further break the arising $\Omega_m-h$ degeneracy by using additional geometrical information from our tomographic analysis, \textit{i.e.}~using multiple redshift bins as the degeneracy direction rotates from a dark energy dominated to a dark matter dominated Universe.

We present a measurement of $H_0$ by combining our present likelihood with the publicly available ACT DR6 CMB-marginalized lensing-only likelihood which includes the \Planck angular power spectrum .\footnote{Publicly available at \url{https://github.com/ACTCollaboration/act_dr6_lenslike}.} In summary, this measurement is sensitive to the turn over scale of the matter power spectrum. Whilst ACT DR6 lensing measures exquisitely well the small scale power spectrum, \Planck provides complimentary information at large scales. Jointly this provides a robust lever arm pinpointing the turnover of the power spectrum. In practice, this corresponds to extending our data vector by the bandpower angular lensing auto spectra paired with the covariance matrix blocks for the auto- and cross-correlation measurements. 

\begin{figure}
    \centering
    \includegraphics[width=1\linewidth]{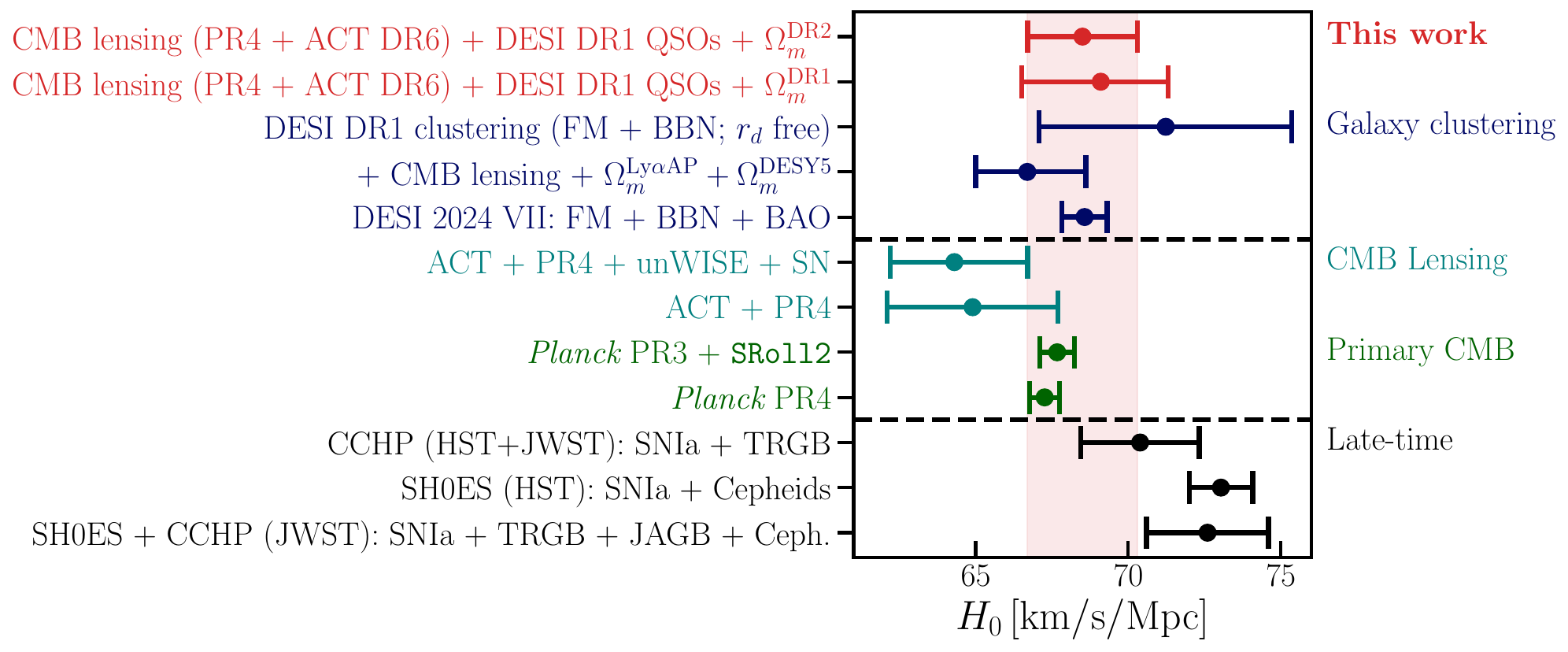}
    \caption{Comparison of constraints on measurements of the Hubble constant from this work (first two rows in red using DR2 and DR1 BAO priors) to other early and late-time probes. The present work performs a sound-horizon free measurement of the Hubble constant by combining CMB lensing auto-spectra from \Planck PR4 and ACT DR6, the cross-correlation of \Planck PR4 lensing maps with the DESI DR1 quasar sample and the auto-spectrum of the quasar sample in the redshift range $0.8\leq z \leq 3.5$. We compare our measurement (red, top row) to DESI 3D galaxy clustering measurements including sound horizon calibrated BAO (BGS + LRG + ELG + QSO) (second row; \cite{Zaborowski:2024wpo}), when combined with \Planck + ACT DR6 CMB lensing and an $\Omega_m$ prior from the \Lya forest (third row; \cite{Zaborowski:2024wpo}) and the DESI full modeling approach \cite{DESI:2024hhd}. Additionally, we show constraints from CMB lensing (ACT DR6, PR4) combined with unWISE galaxy clustering and uncalibrated supernovae from the Pantheon$+$ dataset \cite{Farren:2024rla} (teal) and constraints from the primary CMB (including the low-$\ell$ \texttt{SRoll2} data) \cite{deBelsunce:2021mec,Pagano:2019tci, Rosenberg:2022sdy, Tristram:2023haj}. The bottom panel in black are late-time measurements using the local distance ladder from the Chicago-Carnegie Hubble Program (CCHP; \cite{Freedman:2024eph}) which uses data from the James Webb Space Telescope (JWST) and from the SH0ES team \cite{Riess_2022} which uses data collected by the Hubble Space Telescope (HST). The Tip of the Red Giant Branch calibration method is denoted by TRGB and the J-region Asymptotic Giant Branch method by JAGB, respectively.}
    \label{fig:compilation_H0}
\end{figure}

The data vector consists of adding the band powers for \Planck and ACT DR6. We extend the covariance matrix to include the auto- and cross-spectra of the CMB lensing data with the cross-correlation measurement. Note that as our fiducial, we assume that the correlation between ACT DR6 and $g_i$ equals the correlation of \Planck PR4 and said quasar redshift bin. In addition we sample from a uniform prior on $H_0$ with $\mathcal{U}(40,90)$, and always include BAO data from DESI DR1 or (when specified)  from DR2. Note that BAO constraints lie in the 2-dimensional plane of $\Omega_m$ and $H_0 r_d$, and thus the $\Omega_m$ measurement is automatically independent of $r_d$. We tabulate the results in Table \ref{tab:results} and compare our constraints (we use the DR2 prior as baseline result in the main text) to a compilation of $H_0$ measurements in figure \ref{fig:compilation_H0}. We find excellent agreement with previous works, in particular, with previous DESI galaxy clustering measurements at the $0.60,\, -0.71,\, 0.04\sigma$-level (blue); using CMB lensing alone (ACT DR6 and PR4) we find a $-1.08\sigma$ tension \cite{Madhavacheril2023} which is slightly increased to $-1.46\sigma$ when including information from unWISE galaxy clustering and uncalibrated supernovae from Pantheon$+$ \cite{Farren:2024rla} (teal) and $-0.44,\, -0.66 \sigma$ agreement with the primary CMB (green; the first measurement including a low-$\ell$ \texttt{SRoll2} likelihood). Given our tight BAO prior we obtain tighter constraints than the \textit{Quaia} team \cite{Alonso:2024emk}. The late-time measurement combining data from HST and JWST are in agreement at the $0.71 \sigma$-level with our measurement of the Hubble constant. The SH0ES measurements of $H_0$ is $2.18\sigma$ high compared to our measurement. The combination of SH0ES and CCHP subsamples using JWST agrees at the $1.52 \sigma$ level with our determination of $H_0$.\footnote{We note that the CCHP measurements are based on data from the James Webb Space Telescope (JWST) in contrast to the SH0ES measurement using the Hubble Space Telescope (HST) and JAGB stands for the J-region Asymptotic Giant Branch method.} Whilst most of the constraining power comes from ACT and PR4 paired with the BAO prior on $\Omega_m$ the inclusion of the auto- and cross-correlation improves constraining power by an additional $\sim 4\%$ ($\sim 12\%$)
when using the DR1 (DR2) BAO prior and our fiducial scale cuts. Including the DR2 BAO prior improves the constraining power on $H_0$ by  $\sim$35\%. Increased statistics from DESI year-5 data will be beneficial to explore this further. 

\begin{table}
\centering
\setlength{\tabcolsep}{4pt}
\begin{tabular}{lccc}
\hline\hline \vspace{-2ex} \\ 
Data sets & $H_0$ & $\sigma_8$ &$S_8$ \\[1ex]
\hline \vspace{-2ex} \\ 
$C_{\ell}^{\kappa \kappa,\, \mathrm{PR4+ACT\, DR6}}$ + $\Omega_m^{\rm DR1}$ & $68.8^{+2.3}_{-2.8} \ [68.7]$ & $0.819 \pm 0.018 \ [0.817]$ &$0.810 \pm 0.017 \ [0.814]$\\[1ex]
\qquad $+$ DESI DR1 $\times$ PR4& $69.1^{+2.2}_{-2.6} \ [68.6]$ & $0.823 \pm 0.018 \ [0.817]$ &$0.813 \pm 0.016 \ [0.811]$\\[1ex]
$C_{\ell}^{\kappa \kappa,\, \mathrm{PR4+ACT\, DR6}}$ + $\Omega_m^{\rm DR2}$ & $68.3 \pm 1.9 \ [68.4]$ & $0.816 \pm 0.017 \ [0.818]$ & $0.813 \pm 0.017 \ [0.815]$\\[1ex]
\qquad $+$ DESI DR1 $\times$ PR4& $68.5 \pm 1.8 \ [68.6]$ & $0.819 \pm 0.015 \ [0.822]$ &$0.816 \pm 0.014 \ [0.818]$ \\[1ex]
\hline
\end{tabular}
\caption{Summary of sound-horizon free measurements of $H_0$ together with constraints on $\sigma_8$ and $S_8$ for different datasets and priors. Inclusion of the CMB x DR1 QSO auto- and cross-correlation adds approximately 4\% (12\%) constraining power using $\Omega_m^{\rm DR1}$ ($\Omega_m^{\rm DR2}$) priors from BAO.}
\label{tab:results}
\end{table}

From this analysis we conclude that the high-redshift Universe, as probed by quasars, is consistent with $\Lambda$CDM. We emphasize that the error bars will significantly shrink with the year-five data release of DESI. 

\section{Summary and Conclusion} \label{sec:conclusions}
We present a measurement of the amplitude of matter fluctuations over the redshift range $0.8 \leq z \leq 3.5$ from the cross correlation of 1,223,391 spectroscopically confirmed quasars observed with DESI DR1 and \Planck PR4 CMB lensing maps. We perform a tomographic measurement in three redshift bins with the DESI quasar targets centered at mean redshifts $\bar{z}\sim 1.44$, $2.27$ and $2.75$, which have ample overlap with the CMB lensing kernel. The large lever arm in redshift paired with a BAO prior from galaxy clustering breaks the $\sigma_8-\Omega_m$ degeneracy. The cross-correlation between PR4 lensing maps and DESI DR1 is detected with a signal-to-noise ratio of $27.2$ for the auto-correlation (when excluding shot noise) and $21.7$ for the cross-correlation for the joint analysis of all redshift bins. 

We model the auto- and cross-correlation spectra using a hybrid approach, combining fits to $N$-body simulations and Convolutional Lagrangian Effective Field Theory (CLEFT). Our model consists of three bias parameters $b_1$, $b_2$, $b_s$, magnification bias $s_{\mu}$, the cosmological parameters $\log{A_s}$ and $\Omega_m$ as well as a constant offset to account for shot noise (over which we analytically marginalize). We have performed a suite of consistency checks and shown that our measurement is robust to scale cuts, analysis choices, weighting schemes, fiber assignment effects, redshift errors, the angular integral constraint, and the priors on nuisance and cosmological parameters. 

The auto- and cross-correlation angular power spectra are used to constrain the amplitude of matter fluctuations in the matter-dominated regime to be
\begin{subequations}
\begin{align}
\sigma_8 = 0.929^{+0.059}_{-0.074}\, [0.971] \qquad S_8 = 0.922^{+0.059}_{-0.073}\,[0.969]\,, \qquad &(+\Omega_m^{\rm DR1})\,, \\
\sigma_8 = 0.934^{+0.062}_{-0.082}\, [0.958] \qquad S_8 = 0.930^{+0.061}_{-0.082}\, [0.954]\,, \qquad &(+\Omega_m^{\rm DR2})\,.
\end{align}
\end{subequations}
When extracting constraints from the redshift bins individually, we find much broader yet consistent constraints on the amplitude of matter clustering. This stems from a reduced amount clustering in the $z_2$ and $z_3$ bins compared to the lowest $z_1$ bin. Our constraints largely come from quasi-linear scales ($k\simlt 0.1 \hinvMpc$) for which we explicitly compare our fiducial constraints to constraints from linear theory alone. These are $\sigma_8 = 0.864\pm 0.048$  and $S_8 = 0.859\pm 0.048$ $[0.863]$ (combined with DESI DR1 BAO data) whilst having the drawback that they do not marginalize over higher-order nuisance parameters, magnification bias and shot noise, and are thus more dependent on assumptions about galaxy clustering on small scales; yet yield fully consistent results with predictions from $\Lambda$CDM fits to measurements of the primary CMB from \Planck PR4.

We combine our measurement with the CMB lensing auto-spectrum from the ground-based Atacama Cosmology Telescope (ACT DR6) and Planck PR4 to perform a sound-horizon-free measurement of the Hubble constant yielding 
\begin{subequations}
\begin{align}
    H_0 &= 69.1^{+2.2}_{-2.6} \ [68.6] \,\mathrm{km}\,\mathrm{s}^{-1}\mathrm{Mpc}^{-1} \,, &\sigma_8 = 0.823 \pm 0.018 \ [0.817]\,, \qquad &(+\Omega_m^{\rm DR1})\,, \\
    H_0 &= 68.5 \pm 1.8 \ [68.6] \,\mathrm{km}\,\mathrm{s}^{-1}\mathrm{Mpc}^{-1} \,, &\sigma_8 = 0.819 \pm 0.015 \ [0.822]\,,  \qquad &(+\Omega_m^{\rm DR2})\,,
\end{align}
\end{subequations}
through its sensitivity to the matter-radiation equality scale, $k_{\rm eq}$. Including the auto-and cross-correlation signal tightens the constraints on the Hubble constant by $\sim5$--$10\%$ compared to only using a BAO prior on the energy density and the ACT-PR4 lensing auto-spectrum. 

From our analysis, we do not find any evidence for departure from $\Lambda$CDM from DESI DR1 quasar cross-correlation measurements with \Planck PR4 ($\simlt 2\sigma$ tension which reduces to $\simlt 1\sigma$ when using linear theory). Further, our constraints on $\sigma_8$ are fully consistent with the results obtained from the ``full modeling'' given in table 10 from the DESI DR1 QSO analysis \cite{DESI:2024jis}. We emphasize that we do not find any strong evidence to support the ``$S_8$ tension'' given the current size of the error bars.\footnote{Note that our data is less sensitive to the $S_8$ parameter combination than e.g.~weak lensing experiments.} We leave the inclusion of DESI DR2 quasar data \cite{DESI:2025zgx}, the three-dimensional power spectrum \cite{Maus:2025rvz}, the Lyman-$\alpha$ forest \cite{DESI:2024lzq,deBelsunce:2024knf}, \textit{Quaia} quasar samples \cite{Storey-Fisher:2023gca} or ACT DR6 lensing maps (for the cross-correlation; \cite{Madhavacheril2023}) to future work. This will further help to reduce the uncertainties and stress test $\Lambda$CDM in a barely explored high redshift regime. 

\acknowledgments
RB, EC, SF are supported by Lawrence Berkeley National Laboratory and the Director, Office of Science, Office of High Energy Physics of the U.S. Department of Energy under Contract No. DE-AC02-05CH11231. AK was supported as a CITA National Fellow by the Natural Sciences and Engineering Research Council of Canada (NSERC), funding reference \#DIS-2022-568580.

This material is based upon work supported by the U.S. Department of Energy (DOE), Office of Science, Office of High-Energy Physics, under Contract No. DE–AC02–05CH11231, and by the National Energy Research Scientific Computing Center, a DOE Office of Science User Facility under the same contract. Additional support for DESI was provided by the U.S. National Science Foundation (NSF), Division of Astronomical Sciences under Contract No. AST-0950945 to the NSF's National Optical-Infrared Astronomy Research Laboratory; the Science and Technology Facilities Council of the United Kingdom; the Gordon and Betty Moore Foundation; the Heising-Simons Foundation; the French Alternative Energies and Atomic Energy Commission (CEA); the National Council of Humanities, Science and Technology of Mexico (CONAHCYT); the Ministry of Science, Innovation and Universities of Spain (MICIU/AEI/10.13039/501100011033), and by the DESI Member Institutions: \url{https://www.desi.lbl.gov/collaborating-institutions}.

The DESI Legacy Imaging Surveys consist of three individual and complementary projects: the Dark Energy Camera Legacy Survey (DECaLS), the Beijing-Arizona Sky Survey (BASS), and the Mayall z-band Legacy Survey (MzLS). DECaLS, BASS and MzLS together include data obtained, respectively, at the Blanco telescope, Cerro Tololo Inter-American Observatory, NSF’s NOIRLab; the Bok telescope, Steward Observatory, University of Arizona; and the Mayall telescope, Kitt Peak National Observatory, NOIRLab. NOIRLab is operated by the Association of Universities for Research in Astronomy (AURA) under a cooperative agreement with the National Science Foundation. Pipeline processing and analyses of the data were supported by NOIRLab and the Lawrence Berkeley National Laboratory. Legacy Surveys also uses data products from the Near-Earth Object Wide-field Infrared Survey Explorer (NEOWISE), a project of the Jet Propulsion Laboratory/California Institute of Technology, funded by the National Aeronautics and Space Administration. Legacy Surveys was supported by: the Director, Office of Science, Office of High Energy Physics of the U.S. Department of Energy; the National Energy Research Scientific Computing Center, a DOE Office of Science User Facility; the U.S. National Science Foundation, Division of Astronomical Sciences; the National Astronomical Observatories of China, the Chinese Academy of Sciences and the Chinese National Natural Science Foundation. LBNL is managed by the Regents of the University of California under contract to the U.S. Department of Energy. The complete acknowledgments can be found at \url{https://www.legacysurvey.org/}.

Any opinions, findings, and conclusions or recommendations expressed in this material are those of the author(s) and do not necessarily reflect the views of the U. S. National Science Foundation, the U. S. Department of Energy, or any of the listed funding agencies.

The authors are honored to be permitted to conduct scientific research on I'oligam Du'ag (Kitt Peak), a mountain with particular significance to the Tohono O’odham Nation.

\appendix


\section{Author Affiliations}
\label{sec:affiliations}

\noindent \hangindent=.5cm $^{1}${Lawrence Berkeley National Laboratory, 1 Cyclotron Road, Berkeley, CA 94720, USA}

\noindent \hangindent=.5cm $^{2}${Department of Physics and Astronomy, University of Waterloo, 200 University Ave W, Waterloo, ON N2L 3G1, Canada}

\noindent \hangindent=.5cm $^{3}${Perimeter Institute for Theoretical Physics, 31 Caroline St. North, Waterloo, ON N2L 2Y5, Canada}

\noindent \hangindent=.5cm $^{4}${Waterloo Centre for Astrophysics, University of Waterloo, 200 University Ave W, Waterloo, ON N2L 3G1, Canada}

\noindent \hangindent=.5cm $^{5}${University of California, Berkeley, 110 Sproul Hall \#5800 Berkeley, CA 94720, USA}

\noindent \hangindent=.5cm $^{6}${Institute of Astronomy, University of Cambridge, Madingley Road, Cambridge CB3 0HA, UK}

\noindent \hangindent=.5cm $^{7}${Universit\'{e} Clermont-Auvergne, CNRS, LPCA, 63000 Clermont-Ferrand, France}

\noindent \hangindent=.5cm $^{8}${Department of Physics, Boston University, 590 Commonwealth Avenue, Boston, MA 02215 USA}

\noindent \hangindent=.5cm $^{9}${Dipartimento di Fisica ``Aldo Pontremoli'', Universit\`a degli Studi di Milano, Via Celoria 16, I-20133 Milano, Italy}

\noindent \hangindent=.5cm $^{10}${INAF-Osservatorio Astronomico di Brera, Via Brera 28, 20122 Milano, Italy}

\noindent \hangindent=.5cm $^{11}${Department of Physics \& Astronomy, University College London, Gower Street, London, WC1E 6BT, UK}

\noindent \hangindent=.5cm $^{12}${Instituto de F\'{\i}sica, Universidad Nacional Aut\'{o}noma de M\'{e}xico,  Circuito de la Investigaci\'{o}n Cient\'{\i}fica, Ciudad Universitaria, Cd. de M\'{e}xico  C.~P.~04510,  M\'{e}xico}

\noindent \hangindent=.5cm $^{13}${Department of Astronomy, San Diego State University, 5500 Campanile Drive, San Diego, CA 92182, USA}

\noindent \hangindent=.5cm $^{14}${NSF NOIRLab, 950 N. Cherry Ave., Tucson, AZ 85719, USA}

\noindent \hangindent=.5cm $^{15}${Department of Astronomy \& Astrophysics, University of Toronto, Toronto, ON M5S 3H4, Canada}

\noindent \hangindent=.5cm $^{16}${Department of Physics \& Astronomy and Pittsburgh Particle Physics, Astrophysics, and Cosmology Center (PITT PACC), University of Pittsburgh, 3941 O'Hara Street, Pittsburgh, PA 15260, USA}

\noindent \hangindent=.5cm $^{17}${Institut de F\'{i}sica d’Altes Energies (IFAE), The Barcelona Institute of Science and Technology, Edifici Cn, Campus UAB, 08193, Bellaterra (Barcelona), Spain}

\noindent \hangindent=.5cm $^{18}${Departamento de F\'isica, Universidad de los Andes, Cra. 1 No. 18A-10, Edificio Ip, CP 111711, Bogot\'a, Colombia}

\noindent \hangindent=.5cm $^{19}${Observatorio Astron\'omico, Universidad de los Andes, Cra. 1 No. 18A-10, Edificio H, CP 111711 Bogot\'a, Colombia}

\noindent \hangindent=.5cm $^{20}${Institut d'Estudis Espacials de Catalunya (IEEC), c/ Esteve Terradas 1, Edifici RDIT, Campus PMT-UPC, 08860 Castelldefels, Spain}

\noindent \hangindent=.5cm $^{21}${Institute of Cosmology and Gravitation, University of Portsmouth, Dennis Sciama Building, Portsmouth, PO1 3FX, UK}

\noindent \hangindent=.5cm $^{22}${Institute of Space Sciences, ICE-CSIC, Campus UAB, Carrer de Can Magrans s/n, 08913 Bellaterra, Barcelona, Spain}

\noindent \hangindent=.5cm $^{23}${University of Virginia, Department of Astronomy, Charlottesville, VA 22904, USA}

\noindent \hangindent=.5cm $^{24}${Fermi National Accelerator Laboratory, PO Box 500, Batavia, IL 60510, USA}

\noindent \hangindent=.5cm $^{25}${Institut d'Astrophysique de Paris. 98 bis boulevard Arago. 75014 Paris, France}

\noindent \hangindent=.5cm $^{26}${IRFU, CEA, Universit\'{e} Paris-Saclay, F-91191 Gif-sur-Yvette, France}

\noindent \hangindent=.5cm $^{27}${Center for Cosmology and AstroParticle Physics, The Ohio State University, 191 West Woodruff Avenue, Columbus, OH 43210, USA}

\noindent \hangindent=.5cm $^{28}${Department of Physics, The Ohio State University, 191 West Woodruff Avenue, Columbus, OH 43210, USA}

\noindent \hangindent=.5cm $^{29}${The Ohio State University, Columbus, 43210 OH, USA}

\noindent \hangindent=.5cm $^{30}${Department of Physics, The University of Texas at Dallas, 800 W. Campbell Rd., Richardson, TX 75080, USA}

\noindent \hangindent=.5cm $^{31}${Department of Physics, Southern Methodist University, 3215 Daniel Avenue, Dallas, TX 75275, USA}

\noindent \hangindent=.5cm $^{32}${Department of Physics and Astronomy, University of California, Irvine, 92697, USA}

\noindent \hangindent=.5cm $^{33}${Center for Astrophysics $|$ Harvard \& Smithsonian, 60 Garden Street, Cambridge, MA 02138, USA}

\noindent \hangindent=.5cm $^{34}${Sorbonne Universit\'{e}, CNRS/IN2P3, Laboratoire de Physique Nucl\'{e}aire et de Hautes Energies (LPNHE), FR-75005 Paris, France}

\noindent \hangindent=.5cm $^{35}${Departament de F\'{i}sica, Serra H\'{u}nter, Universitat Aut\`{o}noma de Barcelona, 08193 Bellaterra (Barcelona), Spain}

\noindent \hangindent=.5cm $^{36}${Department of Astronomy, The Ohio State University, 4055 McPherson Laboratory, 140 W 18th Avenue, Columbus, OH 43210, USA}

\noindent \hangindent=.5cm $^{37}${Instituci\'{o} Catalana de Recerca i Estudis Avan\c{c}ats, Passeig de Llu\'{\i}s Companys, 23, 08010 Barcelona, Spain}

\noindent \hangindent=.5cm $^{38}${Departamento de F\'{\i}sica, DCI-Campus Le\'{o}n, Universidad de Guanajuato, Loma del Bosque 103, Le\'{o}n, Guanajuato C.~P.~37150, M\'{e}xico}

\noindent \hangindent=.5cm $^{39}${Instituto Avanzado de Cosmolog\'{\i}a A.~C., San Marcos 11 - Atenas 202. Magdalena Contreras. Ciudad de M\'{e}xico C.~P.~10720, M\'{e}xico}

\noindent \hangindent=.5cm $^{40}${Instituto de Astrof\'{i}sica de Andaluc\'{i}a (CSIC), Glorieta de la Astronom\'{i}a, s/n, E-18008 Granada, Spain}

\noindent \hangindent=.5cm $^{41}${Departament de F\'isica, EEBE, Universitat Polit\`ecnica de Catalunya, c/Eduard Maristany 10, 08930 Barcelona, Spain}

\noindent \hangindent=.5cm $^{42}${Department of Physics and Astronomy, Sejong University, 209 Neungdong-ro, Gwangjin-gu, Seoul 05006, Republic of Korea}

\noindent \hangindent=.5cm $^{43}${CIEMAT, Avenida Complutense 40, E-28040 Madrid, Spain}

\noindent \hangindent=.5cm $^{44}${Department of Physics, University of Michigan, 450 Church Street, Ann Arbor, MI 48109, USA}

\noindent \hangindent=.5cm $^{45}${University of Michigan, 500 S. State Street, Ann Arbor, MI 48109, USA}

\noindent \hangindent=.5cm $^{46}${Department of Physics \& Astronomy, Ohio University, 139 University Terrace, Athens, OH 45701, USA}

\noindent \hangindent=.5cm $^{47}${National Astronomical Observatories, Chinese Academy of Sciences, A20 Datun Road, Chaoyang District, Beijing, 100101, P.~R.~China}



\bibliographystyle{JHEP}
\bibliography{biblio.bib}

\end{document}